\documentclass[preprint2]{aastex}

\newcommand{\EQ}{\begin{equation}}
\newcommand{\EN}{\end{equation}}
\newcommand{\EQA}{\begin{eqnarray}}
\newcommand{\ENA}{\end{eqnarray}}
\newcommand{\eq}[1]{(\ref{#1})}
\newcommand{\Eq}[1]{Eq.~(\ref{#1})}
\newcommand{\Eqs}[2]{Eqs.~(\ref{#1}) and~(\ref{#2})}
\newcommand{\Eqss}[2]{Eqs.~(\ref{#1})--(\ref{#2})}
\newcommand{\Sec}[1]{\S\ref{#1}}
\newcommand{\Fig}[1]{Fig.~\ref{#1}}

\newcommand{\Tab}[1]{Table~\ref{#1}}
\newcommand{\Figs}[2]{Figs~\ref{#1} and \ref{#2}}

\newcommand{\bra}[1]{\langle #1\rangle}
\newcommand{\bbra}[1]{\left\langle #1\right\rangle}

\newcommand{\meanAA}{\overline{\mbox{\boldmath $A$}}}
\newcommand{\meanBB}{\overline{\mbox{\boldmath $B$}}}
\newcommand{\meanJJ}{\overline{\mbox{\boldmath $J$}}}

\newcommand{\meanEE}{\overline{\mbox{\boldmath ${\cal E}$}}}

\newcommand{\eee}{\hat{\mbox{\boldmath $e$}} {}}

\newcommand{\xx}{\mbox{\boldmath $x$} {}}

\newcommand{\zzz}{\mbox{\boldmath $z$} {}}
\newcommand{\uu}{\mbox{\boldmath $u$} {}}

\newcommand{\bb}{\mbox{\boldmath $b$} {}}
\newcommand{\bp}{\mbox{\boldmath $p$} {}}
\newcommand{\qq}{\mbox{\boldmath $q$} {}}
\newcommand{\BB}{\mbox{\boldmath $B$} {}}

\newcommand{\AAA}{\mbox{\boldmath $A$} {}}
\newcommand{\aaa}{\mbox{\boldmath $a$} {}}
\newcommand{\jj}{\mbox{\boldmath $j$} {}}
\newcommand{\JJ}{\mbox{\boldmath $J$} {}}

\newcommand{\ff}{\mbox{\boldmath $f$} {}}

\newcommand{\KK}{\mbox{\boldmath $K$} {}}

\newcommand{\kk}{\mbox{\boldmath $k$} {}}

\newcommand{\nab}{\mbox{\boldmath $\nabla$} {}}

\newcommand{\oo}{\mbox{\boldmath $\omega$} {}}

\newcommand{\emf}{\mbox{\boldmath ${\cal E}$} {}}

\newcommand{\DD}{{\rm D} \, {}}
\newcommand{\dd}{{\rm d} {}}

\newcommand{\ea}{{\em et al. }}

\def\half{{\textstyle{1\over2}}}

\def\onethird{{\textstyle{1\over3}}}

\newcommand{\yr}{\,{\rm yr}}

\newcommand{\yapj}[3]{ #1, {ApJ }{#2}, #3}
\newcommand{\yapjl}[3]{ #1, {ApJ (Letters) }{#2}, #3}

\newcommand{\yan}[3]{ #1, {Astr. Nachr. }{#2}, #3}
\newcommand{\yana}[3]{ #1, {A\&A }{#2}, #3}

\newcommand{\ygafd}[3]{ #1, {Geophys. Astrophys. Fluid Dyn. }{#2}, #3}
\newcommand{\yjfm}[3]{ #1, {J. Fluid Mech. }{#2}, #3}

\newcommand{\ypp}[3]{ #1, {Phys. Plasmas }{#2}, #3}
\newcommand{\yjetp}[3]{ #1, {Sov. Phys. JETP }{#2}, #3}
\newcommand{\yphy}[3]{ #1, {Physica }{#2}, #3}

\newcommand{\yprl}[3]{ #1, {Phys. Rev. Lett. }{#2}, #3}

\newcommand{\ymn}[3]{ #1, {MNRAS }{#2}, #3}
\newcommand{\ynat}[3]{ #1, {Nat }{#2}, #3}

\newcommand{\ypr}[3]{ #1, {Phys. Rev. } {#2}, #3}

\newcommand{\yjgr}[3]{ #1, {JGR }{#2}, #3}

\newcommand{\ybook}[3]{ #1, {#2} (#3)}
\newcommand{\yproc}[5]{ #1, in {#3}, ed. #4 (#5), p.#2}
\newcommand{\pproc}[4]{ #1, in {#2}, ed. #3 (#4), (in press)}

\newcommand{\sgafd}[1]{ #1, {GAFD } (submitted)}

\newcommand{\smn}[1]{ #1, {MNRAS } (submitted)}

\shorttitle{Inverse cascade and nonlinear alpha-effect}
\begin{document}

\title{The inverse cascade and nonlinear alpha-effect in simulations
of isotropic helical hydromagnetic turbulence}

\author{Axel Brandenburg\altaffilmark{1,2}}
\affil{Institute for Theoretical Physics, Kohn Hall,
University of California, Santa Barbara, CA 93106}
\email{brandenb@nordita.dk (\today)}
\altaffiltext{1}{Permanent address: Nordita, Blegdamsvej 17,
DK-2100 Copenhagen \O, Denmark}
\altaffiltext{2}{Also at: Department of Mathematics, University of Newcastle,
Newcastle upon Tyne, NE1 7RU, UK}

\begin{abstract}
A numerical model of isotropic homogeneous turbulence with helical
forcing is investigated. The resulting flow, which is essentially the
prototype of the $\alpha^2$ dynamo of mean-field dynamo theory, produces
strong dynamo action with an additional large scale field on the scale
of the box (at wavenumber $k=1$; forcing is at $k=5$). This large scale
field is nearly force-free and exceeds the equipartition value. As
the magnetic Reynolds number $R_{\rm m}$ increases, the saturation
field strength and the growth rate of the dynamo increase. However,
the time it takes to built up the large scale field from equipartition
to its final super-equipartition value increases with magnetic Reynolds
number. The large scale field generation can be identified as being
due to nonlocal interactions originating from the forcing scale, which
is characteristic of the $\alpha$-effect. Both $\alpha$ and turbulent
magnetic diffusivity $\eta_{\rm t}$ are determined simultaneously using
numerical experiments where the mean-field is modified artificially. Both
quantities are quenched in a $R_{\rm m}$-dependent fashion. The evolution
of the energy of the mean field matches that predicted by an $\alpha^2$
dynamo model with similar $\alpha$ and $\eta_{\rm t}$ quenchings. For
this model an analytic solution is given which matches the results
of the simulations. The simulations are numerically robust in that
the shape of the spectrum at large scales is unchanged when changing
the resolution from $30^3$ to $120^3$ meshpoints, or when increasing
the magnetic Prandtl number (viscosity/magnetic diffusivity) from 1
to 100. Increasing the forcing wavenumber to 30 (i.e.\ increasing the
scale separation) makes the inverse cascade effect more pronounced,
although it remains otherwise qualitatively unchanged.
\end{abstract}


\section{Introduction}

The generation of large scale magnetic fields from small scale turbulence
is important in many astrophysical bodies (planets, stars, accretion discs
and galaxies).  Over many decades the $\alpha\omega$-dynamo concept has
been invoked to explain large scale magnetic field generation (Moffatt
1978, Parker 1979, Krause \& R\"adler 1980). Over recent years however
numerical simulations have become available that produce large scale
fields with appreciable magnetic energy, sometimes even exceeding the
turbulent kinetic energy (e.g.\ Glatzmaier \& Roberts 1995, Brandenburg
\ea 1995, Ziegler \& R\"udiger 2000). Whether or not large scale field
generation to such amplitudes is related to the $\alpha$-effect remains
debatable (e.g.\ Cattaneo \& Hughes 1996, Brandenburg \& Donner 1997).

The $\alpha$-effect is a key ingredient to many astrophysical dynamo models.
The purpose of this paper is, therefore, to study a simple system that is
prototypical of the $\alpha$-effect: homogeneous isotropic turbulence
that lacks mirror symmetry. [Astrophysical dynamos often work in conjunction
with shear, i.e.\ the $\omega$-effect: this case is studied in
a second paper (Brandenburg, Bigazzi, \& Subramanian 2000)]. An isotropic
helical turbulent flow is accomplished by adopting a body
force corresponding to plane polarized waves in random directions (but
constant polarization) with wavelengths short compared with the size
of the box. Since the seminal papers by Frisch \ea (1975) and Pouquet,
Frisch, \& L\'eorat (1976) we know that there should be an inverse
cascade of magnetic helicity, which has also been demonstrated using
direct numerical simulations (e.g.\ Meneguzzi, Frisch, \& Pouquet 1981,
Balsara \& Pouquet 1999). However, to our knowledge there has never been
a detailed study of the spatial magnetic field patterns obtained from
the inverse cascade, nor has there been a quantitative identification of
the classical $\alpha$-effect in mean-field dynamo theory. Furthermore,
the Reynolds and Prandtl number dependences of this process have not been
fully explored yet. In the present paper we study models with strongly
helical forcing at different Reynolds numbers. We also investigate some
models where the magnetic Prandtl number (viscosity/magnetic diffusivity)
is increased from 1 to 100. This may be important in connection with the
galactic magnetic field, and there are some serious concerns that the
inverse cascade may not be efficient at large magnetic Prandtl numbers.

\section{The model}

We consider a compressible isothermal gas with constant sound speed
$c_{\rm s}$, constant dynamical viscosity $\mu$, constant magnetic
diffusivity $\eta$, and constant magnetic permeability $\mu_0$. The
governing equations for density $\rho$, velocity $\uu$, and magnetic
vector potential $\AAA$, are given by
\EQ
{\DD\ln\rho\over\DD t}=-\nab\cdot\uu,
\label{dlnrhodt}
\EN
\EQ
{\DD\uu\over\DD t}=-c_{\rm s}^2\nab\ln\rho+{\JJ\times\BB\over\rho}
+{\mu\over\rho}(\nabla^2\uu+\onethird\nab\nab\cdot\uu)+\ff,
\label{dudt}
\EN
\EQ
{\partial\AAA\over\partial t}=\uu\times\BB-\eta\mu_0\JJ,
\label{dAdt}
\EN
where ${\rm D}/{\rm D}t=\partial/\partial t+\uu\cdot\nab$ is the
advective derivative, $\BB=\nab\times\AAA$ is the magnetic field,
$\JJ=\nab\times\BB/\mu_0$ is the current density, and $\ff$ is a
random forcing function.

We use periodic boundary conditions in all three directions
for all variables. This implies that the mass in the box
is conserved, i.e.\ $\bra\rho=\rho_0$, where $\rho_0$ is the value
of the initially uniform density, and angular brackets denote
volume averages. We adopt a forcing function $\ff$ of the form
\EQ
\ff(\xx,t)=\mbox{Re}\{N\ff_{\kk(t)}\exp[i\kk(t)\cdot\xx+i\phi(t)]\},
\EN
where $\kk(t)=(k_x,k_y,k_z)$ is a time dependent wavevector,
$\xx=(x,y,z)$ is position, and $\phi(t)$ with $|\phi|<\pi$ is
a random phase. On dimensional grounds the normalization factor
is chosen to be
$N=f_0 c_{\rm s}(kc_{\rm s}/\delta t)^{1/2}$, where $f_0$ is a
nondimensional factor, $k=|\kk|$, and $\delta t$ is the length of the
timestep. We focus on the case where $|\kk|$ is around $k_{\rm f}=5$,
and select at each timestep randomly one of the 350 possible vectors
in $4.5<|\kk|<5.5$. We force the system with eigenfunctions of the
curl operator,
\EQ
\ff_{\kk}={\kk\times(\kk\times\eee)-i|\kk|(\kk\times\eee)
\over2\kk^2\sqrt{1-(\kk\cdot\eee)^2/\kk^2}},
\EN
where $\eee$ is an arbitrary unit vector needed in order
to generate a vector $\kk\times\eee$ that is perpendicular
to $\kk$. Note that $|\ff_{\kk}|^2=1$ and, in particular,
$i\kk\times\ff_{\kk}=|\kk|\ff_{\kk}$, so the helicity density of this
forcing function satisfies
\EQ
\ff\cdot\nab\times\ff=|\kk|\ff^2>0
\EN
at each point in space. We note that since the forcing function is like
a delta-function in $\kk$-space, this means that all points of $\ff$
are correlated at any instant in time, but are different at the next
timestep. Thus, the forcing function is delta-correlated in time (but
the velocity is not).

We adopt nondimensional quantities by measuring $\uu$ in units of $c_{\rm
s}$, $\xx$ in units of $1/k_1$, where $k_1$ is the smallest wave number
in the box, which has a size of $L=2\pi$, density in units of $\rho_0$,
and $\BB$ is measured in units of $\sqrt{\mu_0\rho_0}\,c_{\rm s}$. This
is equivalent to putting
\EQ
c_{\rm s}=k_1=\rho_0=\mu_0=1.
\EN
In the following we always quote the {\it mean} kinematic viscosity
$\nu\equiv\mu/\rho_0$, which is close to the actual kinematic viscosity
$\mu/\rho$ because the Mach numbers considered in the present paper are
less than one.

We advance the equations in time using a third order Runge-Kutta scheme
and sixth order explicit centered derivatives in space. In all cases
presented we chose $f_0=0.1$, which yields rms Mach numbers around
0.1--0.3, and peak values less than one.

Our initial condition is $\ln\rho=\uu=0$, and $\AAA$ is a smoothed
gaussian random field that is delta-correlated in space, so the initial
magnetic energy spectrum is $E_{\rm M}(k)\sim k^4$ with a decline at high
wavenumbers.

\section{Results}

All the runs are summarized in \Tab{T1}. The definition of various
entries to the table are given below, together with an outline of the
general behavior of the solutions.

After about 30 time units the rms velocity, $u_{\rm rms}$, reaches
an approximate equilibrium amplitude of up to 0.3. (Since $c_{\rm
s}=1$, this is also the Mach number.) This velocity corresponds to a
turnover time of $\tau=\ell_{\rm f}/u_{\rm rms}\approx4$ time units,
where $\ell_{\rm f}=2\pi/k_{\rm f}$ is the forcing scale. We note
that the value of $\tau$ is approximately equal to the value of the
correlation time obtained from the temporal correlation function of the
velocity. The value of $u_{\rm rms}$ is somewhat smaller for smaller
Reynolds number. The flow has strong positive helicity, as measured by
the relative helicity $\bra{\oo\cdot\uu}/(\omega_{\rm rms}u_{\rm rms})$,
which can be as large as 70\% (or even larger when the Reynolds number
is smaller). Here, $\oo=\nab\times\uu$ is the vorticity.

The growth rate of the magnetic field is determined
as $\lambda=\dd\ln\bra{\BB^2}^{1/2}_{\rm lin}/\dd t$, where the
subscript `lin' refers to early times when the field is still weak
on all scales. A hyphen in the table indicates that the run has been
restarted from another run, so no data are available for the linear
growth phase. Also given is the growth rate normalized with the turnover,
$\tau=\ell_{\rm f}/\bra{\uu^2}^{1/2}_{\rm lin}$.

In order to assess the Reynolds number dependence of our results we have
performed three runs with $\nu=\eta=0.002$ (Run~1), $\nu=\eta=0.005$
(Run~2), and $\nu=\eta=0.01$ (Run~3); see \Tab{T1} for a summary. In order
to assess the dependence on magnetic Prandtl number we have additional
runs with $\nu/\eta=20$ (Run~4) and 100 (Run~5), as well as one with
$\nu/\eta=0.1$ (Run~7). These runs will be explained in detail in
\Sec{Sseparation}. In Runs~1--5 and 7 the forcing wavenumber was around
5, but in Run~6 we increased it to 30 in order to study the properties
of larger scale separation; see \Sec{Sseparation}. The root-mean-square
values of various quantities are reasonably well converged, as can
be gauged by comparing Run~2 ($60^3$ meshpoints) with Run~2l ($30^3$
meshpoints), which has the same values of $\eta$ and $\nu$. We return to
a detailed discussion on the Reynolds number dependence in \Sec{SRe_dep}.

In the table we give various magnetic Reynolds numbers: $R_{\rm m}$ is
based on the box size ($=2\pi$) and the velocity $\bra{\uu^2}^{1/2}_{\rm
sat}$ at the time when the magnetic field is saturated, $R_{\rm
m,lin}$ is the same but during the linear growth phase (using
$\bra{\uu^2}^{1/2}_{\rm lin}$), $R_{\rm m,\lambda}$ is based on the
Taylor microscale $\bra{\uu^2}^{1/2}_{\rm lin}/\bra{\oo^2}^{1/2}_{\rm
lin}$ and $R_{\rm m,forc}$ is based on the forcing scale $\ell_{\rm f}$
and $\bra{\uu^2}^{1/2}_{\rm lin}$. The critical values of $R_{\rm m,forc}$
for the onset of dynamo action are also given and are typically between 7
and 9. In all cases the onset for dynamo action occurs for $\nu/\eta<1$,
i.e.\ for magnetic Prandtl numbers less than unity.

\begin{deluxetable}{lccccccccc}
\tabletypesize{\scriptsize}
\tablecaption{Summary of runs.
\label{T1}}
\tablewidth{0pt}
\tablehead{
    & \colhead{Run~1} & \colhead{Run~2l} & \colhead{Run~2} & \colhead{Run~3} 
& \colhead{Run~3p} & \colhead{Run~4} & \colhead{Run~5} & \colhead{Run~6} & \colhead{Run~7} \\
}
\startdata
mesh points                  &$30^3$&$30^3$&$60^3$&$120^3$&$120^3$&$120^3$&$120^3$&$120^3$&$120^3$\\
$\nu$                        & 0.01 & 0.005& 0.005& 0.002 & 0.002&  0.02 &  0.02 & 0.001  & 0.002 \\
$\nu/\eta$                   &   1  &   1  &    1 &    1  &    2 &   20  &  100  &    1   &  0.1  \\
$k_{\rm f}$                  &   5  &   5  &    5 &    5  &    5 &    5  &    5  &   30   &   5   \\
$\bra{\uu^2}^{1/2}_{\rm lin}$& 0.16 & 0.23 & 0.22 & 0.29  &   -- &  0.11 &  0.114&  0.082 &  0.29 \\
$\bra{\uu^2}^{1/2}_{\rm sat}$& 0.12 & 0.15 & 0.15 & 0.18  & 0.19 &  0.10 &  0.104&  0.062 &  0.20 \\
$R_{\rm m}$                  &  80  &  200 &  200 &  600  & 1200 &   700 &  3300 &   400  &    60 \\
$R_{\rm m,lin}$              & 100  &  300 &  300 &  900  & 1800 &   700 &  3600 &   500  &    90 \\
$R_{\rm m,\lambda}$          &   3  &    9 &    9 &   23  &   46 &    21 &   112 &     3  &     2 \\
$R_{\rm m,forc}$             &  20  &   60 &   60 &  180  &  360 &   140 &   700 &    17  &    18 \\
$R_{\rm m,forc,crit}$        &  7.3 &   -- &  6.9 &  8.9  &  8.9 &    12 &    12 &    --  &   8.9 \\
$\lambda$                    & 0.026& 0.06 & 0.056& 0.067 &  --  &  0.03 &   0.04&  0.075 &  0.036\\
$\lambda\tau$                & 0.24 & 0.32 & 0.34 & 0.30  &  --  &  0.34 &  0.48 &  0.19  &  0.16 \\
$\bra{\BB^2}^{1/2}_{\rm sat}$& 0.18 & 0.27 & 0.28 & 0.38  & 0.40 & $>0.2$&$>0.21$& $>0.2$ &  0.18 \\
$\bra{\JJ^2}^{1/2}_{\rm sat}$& 0.44 & 0.75 & 0.76 & 1.27  & 1.56 &  0.7  &  1.05 &   1.5  &  0.46 \\
$\bra{\oo^2}^{1/2}_{\rm lin}$& 0.80 & 1.12 & 1.12 & 1.81  & 1.81 &  0.58 &  0.58 &   2.4  &  1.76 \\
$\bra{\oo^2}^{1/2}_{\rm sat}$& 0.65 & 0.78 & 0.82 & 1.23  & 1.38 &  0.55 &  0.55 &   1.8  &  1.03 \\
$\bra{\oo\cdot\uu}_{\rm lin}$& 0.13 & 0.24 & 0.24 & 0.37  & 0.37 &  0.063 & 0.063&   0.20 &  0.36 \\
$\bra{\oo\cdot\uu}_{\rm sat}$& 0.08 & 0.10 & 0.11 & 0.16  & 0.17 &  0.058 & 0.055&   0.10 &  0.19 \\
$\bra{\JJ\cdot\BB}_{\rm max}$& 0.007& 0.025& 0.027& 0.07 &$>0.03$&  0.025 &$>0.06$& 0.040 & 0.006 \\
$\bra{\JJ\cdot\BB}_{\rm max}/(J_{\rm rms}B_{\rm rms})$
                             & 0.22 & 0.20 & 0.20 & 0.25  &  --  &  0.23  & 0.18 &  0.25  &  0.17 \\
$|\bra{\JJ\cdot\BB}/\bra{\BB^2}|_{\rm lin}$
                               &0.9 & 1.3  & 1.1  &1--1.5 &  --  &  2.2   & 2.4  &   5.5  &  0.70 \\
\enddata
\end{deluxetable}

\begin{figure}[h!]\plotone{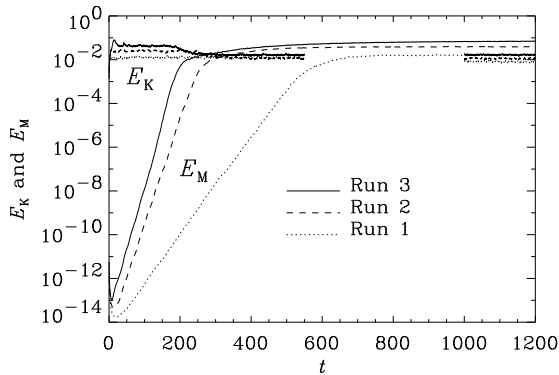}\caption[]{
Evolution of magnetic energy and kinetic energy (per volume) for three
values of the magnetic Reynolds number, $R_{\rm m}$. Note that
the growth rate and the saturation level increase with increasing $R_{\rm
m}$. For reasons of clarity the curves of kinetic energy are not shown
in the range $550<t<1000$.
}\label{Fpenerg}\end{figure}

The evolution of magnetic and kinetic energies ($E_{\rm M}$ and $E_{\rm
K}$) is shown in \Fig{Fpenerg}. Note that $E_{\rm K}$ decreases after
$E_{\rm M}$ has reached its saturation value. (We note that even
after saturation the field continues to grow somewhat, but this will
be discussed in full detail in \Sec{SRe_dep}.) The relative kinetic
helicity changes only slightly before and after saturation. Both the
growth rate and the saturation level of the magnetic field increase
with increasing Reynolds number, and are likely to reach some asymptotic
value at sufficiently large Reynolds number.

The level of turbulence may be characterized by the ratio of the turbulent
to the microscopic diffusion coefficient for a passive scalar, $D_{\rm
t}/D$. The standard estimate is $D_{\rm t}={1\over3}u_{\rm rms}\ell$,
so $D_{\rm t}/D={1\over3}\mbox{Re}_{\rm forc}$. For Run~3 we have
$\mbox{Re}_{\rm forc}=180$, so we expect $D_{\rm t}/D=60$. The actual
value obtained by solving the passive scalar advection-diffusion
equation simultaneously with \Eqss{dlnrhodt}{dAdt} is somewhat
smaller; see \Sec{mf_interpret} where we find $D_{\rm t}/D\approx40$ for
weak fields. This is probably due to the absence of a proper inertial
range. Ideally one would like to simulate higher levels of turbulence,
which requires higher resolution. Certain questions can therefore not be
addressed in a satisfactory manner, for example what are the spectral
properties of the magnetic field, especially at large magnetic Prandtl
numbers. Addressing this requires the presence of a sufficiently extended
inertial range. Other aspects may very well be addressed, for example
what is the behavior of the large scale field and how does it depend on
Reynolds and Prandtl numbers. We shall show that the spectral properties
are well converged at large scales, but the time scales for reaching a
final state increase with magnetic Reynolds number. In order to address
these questions it is important that there is sufficient scale separation
between the energy carrying scale and the scale of the box. Furthermore
it is important to allow for sufficient separation between dynamic
and resistive timescales in order to identify properly the mechanisms
affecting large scale dynamo action. A factor of 5 in scale separation
seems to be a good compromise allowing still some degree of turbulent
mixing to take place.

\subsection{The inverse cascade}

\begin{figure}[h!]\plotone{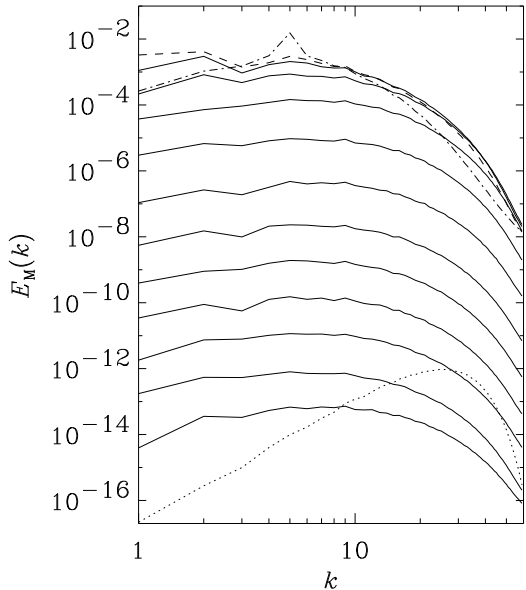}\caption[]{
Spectra of magnetic energy for Run~3 during the initial growth phase at
$t=0$ (dotted line), $t=20$, 40, ..., 220 (solid lines) and
$t=240$ (dashed line). The kinetic energy spectrum (time averaged between
$600\leq t\leq1000$) is shown for comparison (dash-dotted line give the
kinetic energy spectrum.
}\label{Fpspec_growth}\end{figure}

Consistent with previous studies in this field (e.g.\ Meneguzzi \ea 1981,
Balsara \& Pouquet 1999), we find the development of large scale fields
through an inverse cascade effect of the magnetic helicity. This is best
seen in the evolution of magnetic energy spectra, $E_{\rm M}(k)$; see
\Fig{Fpspec_growth}. The kinetic energy spectrum, $E_{\rm K}(k)$, is
also shown.

The random initial condition has a $k^4$ powerspectrum, corresponding to
a delta-correlated vector potential. However, even though the initial
field was smoothed, the spectrum is deformed significantly during the
first few timesteps. During the interval $20\leq t\leq200$ the spectrum
is nearly shape invariant and grows at all scales at the same rate;
see \Fig{Fpspec_growth}. This is typical of small-scale dynamos (Kazantsev 1968).

\begin{figure}[h!]\plotone{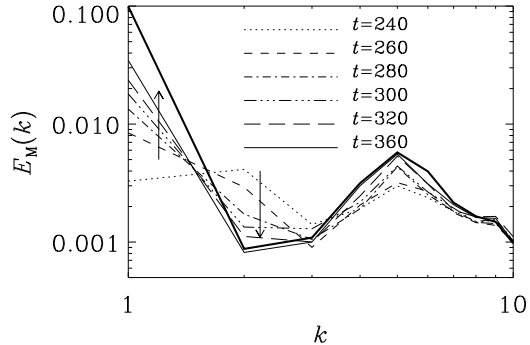}\caption[]{
Spectra of magnetic energy for Run~3 during the saturation phase at times
between $t=240$ and 360. The time averaged spectrum for times between
600 and 1000 is shown as a thick line. (Only the range $1\leq k\leq10$
is shown.)
}\label{Fpspec_growth2}\end{figure}

At $t=200$ the magnetic energy approaches equipartition with the kinetic
energy at small scales. After $t=240$ the magnetic energy is in slight
superequipartition with the kinetic energy at $k>10$. This marks the
beginning of a more complicated process (\Fig{Fpspec_growth2}) during
which the field at the largest possible scale ($k=1$) continues to grow,
but the field at intermediate wavenumbers ($k=2$, 3, and 4) begins to
decline. This process is essentially completed by the time $t=400$.
The significance of this process becomes clear when looking at the
magnetic field evolution in real space.

\subsection{The emergence of a large scale field}

\begin{figure}[h!]\plotone{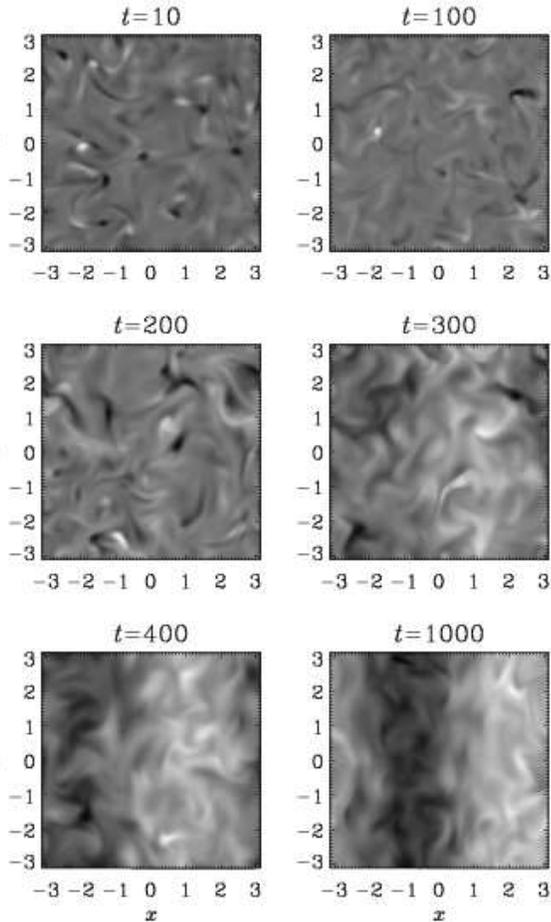}\caption[]{
Gray-scale images of cross-sections of $B_x(x,y,0)$ for Run~3 at
different times showing the gradual build-up of the large scale magnetic
field after $t=300$. Dark (light) corresponds to negative (positive)
values. Each image is scaled with respect to its min and max values.
}\label{Fpimages}\end{figure}

\begin{figure}[h!]\plotone{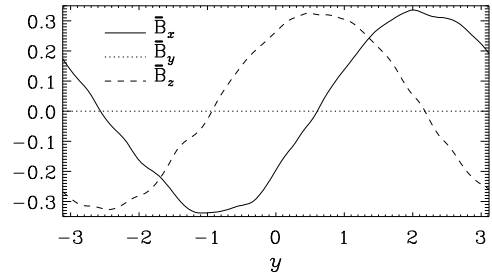}\caption[]{
The three magnetic field components averaged in the $x$ and $z$ directions
at $t=1000$. Note the $90^\circ$ phase shift between $\bra{{\overline B}_x}(y)$ and
$\bra{{\overline B}_z}(y)$, and that the functional form is nearly sinusoidal. Run~3.
}\label{Frmean1000}\end{figure}

Although the magnetic field has reached equipartition already at
$t\approx200$, and its scale began to reach the largest possible scale
of the box, it took another 100 time units for the large scale field
at scale $k=1$ to fully develop and, more importantly, to suppress the
power at intermediate scales. Looking at $xy$ and $yz$ cross-sections, two
components of the field ($B_x$ and $B_z$) show the development of a large
scale sinusoidal modulation through the entire box. In \Fig{Fpimages} we
show $xy$ slices of $B_x$, but the $yz$ cross-sections look qualitatively
similar, except for a $90^\circ$ phase shift of $B_z$ in the $y$
direction. This systematic phase shift is seen more clearly in a plot
of the three field components averaged in the $x$ and $z$ directions;
see \Fig{Frmean1000}.

\begin{figure}[h!]\plotone{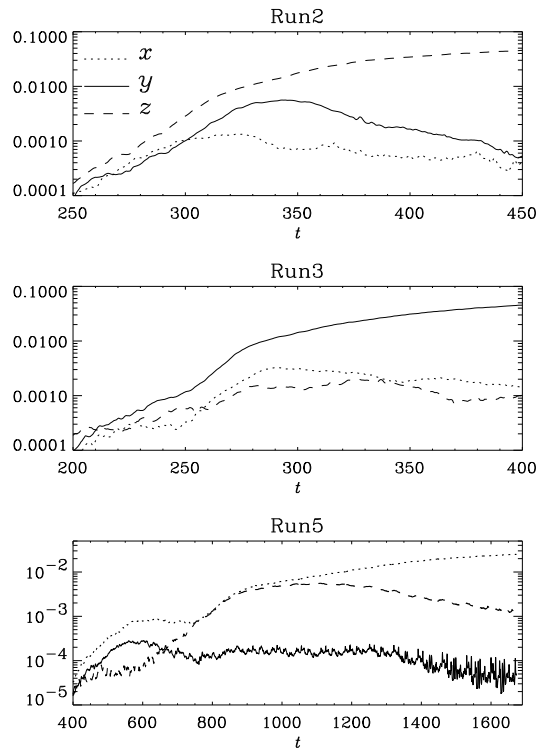}\caption[]{
Magnetic energies (per volume) of those components of the large scale
field whose wave vectors point in the $x$, $y$, or $z$ direction. Which
of the three possible force-free solutions is selected in the end depends
on chance.
}\label{Fpbmean}\end{figure}

Although our forcing is isotropic, one particular direction has
been selected by the large scale magnetic field. In Runs~1 and 3 it
was the $y$-direction, in Run~2 the $z$-direction, and in Run~5 the
$x$-direction. Which direction is selected depends on fine details
of the initially random condition. Nevertheless, it is not until the
time of saturation that the final selection is established, as can be
seen in \Fig{Fpbmean} where we plot the magnetic energies of the mean
field for the three possible directions, denoted by $E(K_x)$, $E(K_y)$,
$E(K_z)$. So,
\EQ
E(K_x)=\bra{\bra{\BB}_{yz}^2}_x,
\EN
\EQ
E(K_y)=\bra{\bra{\BB}_{xz}^2}_y,
\EN
\EQ
E(K_z)=\bra{\bra{\BB}_{xy}^2}_z,
\EN
where the subscripts denote the directions of averaging. Thus, for any
direction $\KK$, say the $x$-direction, we define corresponding mean
fields by averaging in the two perpendicular directions ($y$ and $z$
in this case), and then we calculate their mean squared value. The time
of selection, i.e.\ when one of the three $E(K_i)$ becomes dominant, is
earlier in the large Reynolds number cases.

\begin{figure}[h!]\plotone{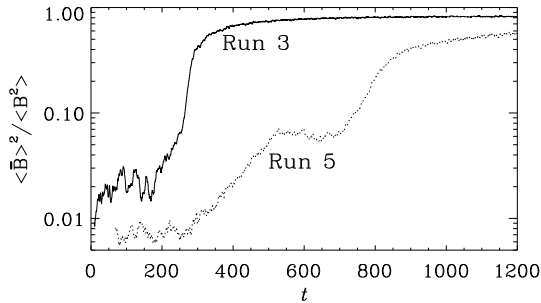}\caption[]{
Evolution of the ratio $\bra{\meanBB}^2/\bra{\BB^2}$,
for Runs~3 and 5. Note that strong large scale fields are
obtained even for large magnetic Reynolds and Prandtl numbers.
}\label{Fpq}\end{figure}

A quantity of theoretical interest is the ratio
$\bra{\meanBB^2}/\bra{\BB^2}$, which characterizes the fraction of space
occupied by the large scale field. Initially this ratio is just $\sim2\%$
(for Run~3) and $\sim0.7\%$ (for Run~5), but later it begins to level
off near 80\% (\Fig{Fpq}). Most likely real astrophysical dynamos
are far less effective in producing such clean large scale fields,
because in reality the helicity of the effective forcing will be far
less than 100\%.  Nevertheless, it is important to notice that it is at
least theoretically possible to achieve large scale field energies near
or in excess of the kinetic energy, even though the magnetic Reynolds
number is reasonably high.

We note that the phase of the large scale field may be drifting slowly
as long as the large scale magnetic energy has not yet reached a fully
steady state. In Run~3, for example, the phase was still drifting slowly
in the $y$ direction (speed $\sim+1.5\eta k_1$), but then it began to
settle after $t\approx1000$.

\subsection{Spectral helicity and energy transfer}

\begin{figure}[h!]\plotone{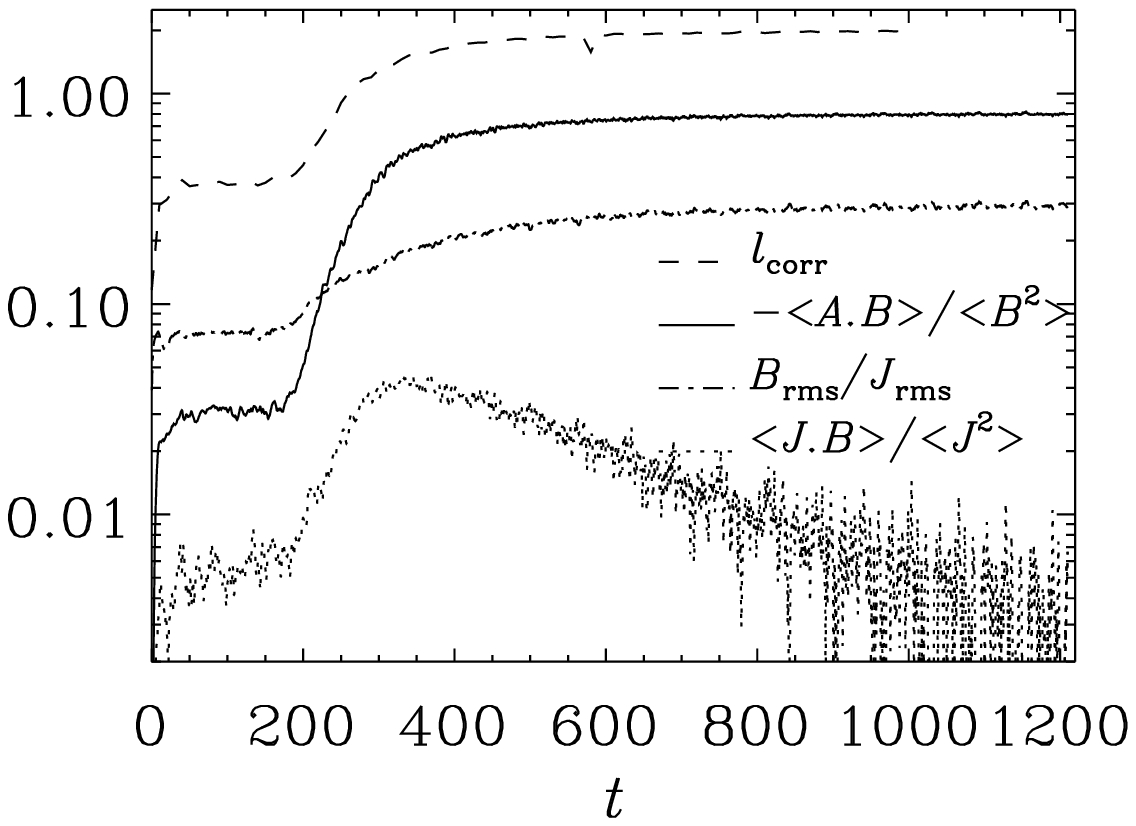}\caption[]{
Evolution of $-\bra{\AAA\cdot\BB}/\bra{\BB^2}$, compared with other
magnetic length scales. Run~3.
}\label{Fpcorrel}\end{figure}

\begin{figure}[h!]\plotone{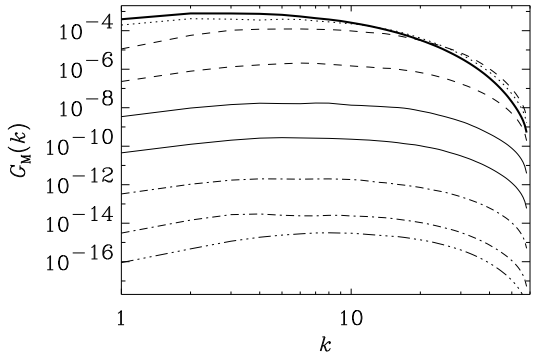}\caption[]{
Spectra of magnetic helicity flux at $t=10$, 30, and 60
(dot-dashed), 100 an 130 (solid), 160 and 200 (dashed), 300 (dotted),
and for the time average between $t=600$ and 1000 (thick line).
}\label{Fpspec_helic2}\end{figure}

\begin{figure}[h!]\plotone{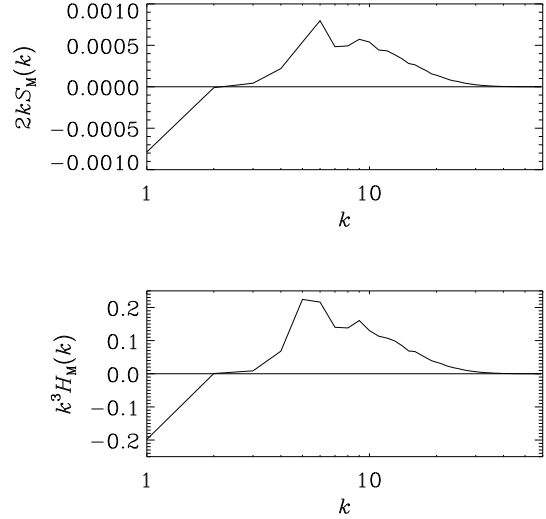}\caption[]{
Divergence of magnetic helicity flux, $S_{\rm M}(k)$, scaled by $k$,
so the area under the positive and negative parts of the curve are
the same when plotted in a lin-log plot. Because of resistive losses
of positive magnetic helicity at small scales, the resulting magnetic
helicity, $H_{\rm M}(k)$, is now dominated by negative helicity at
large scales.  The lower plot shows $H_{\rm M}(k)$, scaled by $k^3$ to
show the contributions from small scales. Both spectra are time averages
over the interval $600\leq t\leq1000$.
}\label{Fpspec_helic}\end{figure}

The primary reason for the large scale field generation is related to
magnetic helicity conservation. Once helicity is injected into the
system, it tends to make the magnetic field also helical, as is seen from
\Fig{Fpcorrel}. For a closed or periodic system however the net magnetic
helicity is conserved, except for diffusion at small scales, i.e.\
\EQ
{\dd\over\dd t}\bra{\AAA\cdot\BB}=-2\eta\bra{\JJ\cdot\BB}.
\label{HelCons}
\EN
Thus, if the magnetic field is to become helical, it must at first
have equal amounts of positive and negative helicity. This feature,
which is familiar in MHD (e.g.\ Seehafer 1996, Ji 1999), is also seen
in hydrodynamical simulations (Biferale \& Kerr 1995). At later times,
however, magnetic diffusion can destroy magnetic helicity at small scales,
leaving magnetic helicity of opposite sign at large scales. This is best
described by the evolution equation of the magnetic helicity spectrum
which can be derived from the Fourier transformed induction equation
\eq{dAdt},
\EQ
{\partial\hat{\AAA}_{\kk}\over\partial t}=\hat{\emf}_{\kk}-\eta\hat{\JJ}_{\kk},
\label{dAkdt}
\EN
where hats and subscripts $\kk$ indicate three-dimensional Fourier
transformation, and $\emf=\uu\times\BB$ is the electromotive
force. We write down the corresponding equation for the
evolution of $\hat{\BB}_{\kk}=i\kk\times\hat{\AAA}_{\kk}$
and derive from these equations the evolution equation for
$\hat{\AAA}_{\kk}\cdot\hat{\BB}_{\kk}^*$, where asterisks denote
complex conjugation. Note that this is gauge invariant, because
adding a gradient to $\hat{\AAA}_{\kk}$ corresponds to adding
an $i\kk\cdot\hat{\BB}_{\kk}^*$ term which vanishes, because
the magnetic field is solenoidal. We denote the real parts of the
shell-integrated spectra of $\hat{\AAA}_{\kk}\cdot\hat{\BB}_{\kk}^*$
and $\hat{\emf}_{\kk}\cdot\hat{\BB}_{\kk}^*$ by $H_{\rm M}(k,t)$ and
$S_{\rm M}(k,t)$, respectively, and obtain
\EQ
{\partial\over\partial t}H_{\rm M}(k,t)=2S_{\rm M}(k,t)-2\eta k^2H_{\rm M}(k,t).
\label{hel_evol}
\EN
Note that $\int H_{\rm M}(k,t)dk=\bra{\AAA\cdot\BB}$ and, because of
helicity conservation, $\int S_{\rm M}(k,t)dk=0$, so it makes sense to
write \Eq{hel_evol} in the form
\EQ
{\partial\over\partial t}H_{\rm M}(k,t)=-
{\partial\over\partial k}G_{\rm M}(k,t)-2\eta k^2H_{\rm M}(k,t),
\EN
where we have defined the spectral flux of helicity,
\EQ
G_{\rm M}(k,t)=\int_k^{\infty}2S_{\rm M}(k',t)\,\dd k',
\EN
which is plotted in \Fig{Fpspec_helic2} for different times. The magnetic
helicity flux, $G_{\rm M}(k,t)$, is found to be always positive and
its peak shifts from small scales ($k\approx10$) at early times to
large scales ($k=2-3$) at later times when the magnetic field becomes
dynamically important. Positive magnetic helicity is being produced on
the right of the maximum of $G_{\rm M}(k,t)$ and negative on the left;
see \Fig{Fpspec_helic}.

\begin{figure}[h!]\plotone{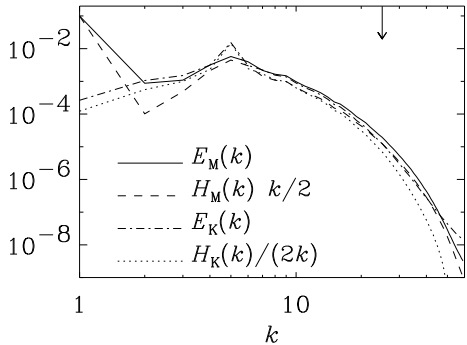}\caption[]{
Time-averaged spectra of kinetic and magnetic energy, as well as kinetic
and magnetic helicity. Note that the magnetic energy exceeds the kinetic
energy at $k=1$, and that the inequality \eq{realizability} is almost
saturated, except near $k=2$ and 3. The corresponding realizability
condition for the kinetic helicity, on the other hand, is not very sharp.
The dissipative cutoff wavenumber, $\bra{\oo^2/\nu^2}^{1/4}$, is indicated
by an arrow at the top.
}\label{Fpspec_all}\end{figure}

In view of the realizability condition,
\EQ
E_{\rm M}(k,t)\geq \half k H_{\rm M}(k,t)
\label{realizability}
\EN
(e.g.\ Moffatt 1978), the spectral magnetic helicity can be viewed as
the {\it driver} of spectral magnetic energy: while small scale magnetic
helicity is being destroyed, an equal amount gets into the large
scales, and this must necessarily enhance the magnetic energy so
as to satisfy \eq{realizability}. Indeed, in the present simulations
the inequality \eq{realizability} is almost saturated at all scales,
except at intermediate scales $2\leq k\leq4$; see \Fig{Fpspec_all}.

In order to determine the dominant interactions leading to the generation
of the large scale field at $k=1$ we now consider the spectral energy
equation,
\EQ
{\partial\over\partial t}E_{\rm M}(k,t)=2T_{\rm M}(k,t)-2\eta k^2E_{\rm M}(k,t),
\EN
where the transfer function of magnetic energy, $T_{\rm M}(k,t)$, is the
shell-integrated spectrum of the real part of
$\hat{\emf}_{\kk}\cdot\hat{\JJ}_{\kk}^*$. Since $\emf=\uu\times\BB$, this
corresponds really to a triple product,
\EQ
\sum_{\kk=\bp+\qq}(\hat{\uu}_{\bp}\times\hat{\BB}_{\qq})\cdot\hat{\JJ}_{\kk}^*,
\EN
where the skew product can also be written as
$-\hat{\uu}_{\bp}\cdot(\hat{\JJ}_{\kk}^*\times\hat{\BB}_{\qq})$, emphasizing
that this term corresponds to the work done against the Lorentz force.
In order to identify the dominant interactions we have calculated,
in real space, the spectral transfer matrix
\EQ
T_{\rm M}(k,p,q,t)=-\bbra{\uu_p\cdot(\JJ_k\times\BB_q)},
\EN
where angular brackets denote volume averages and the subscripts $k$,
$p$, and $q$ denote Fourier filtering around the corresponding wavenumber
(by $\pm1/2$). (In this notation $\bra{\AAA_k\cdot\BB_k}$, for example,
is exactly the same as the helicity spectrum.)

\begin{figure}[h!]\plotone{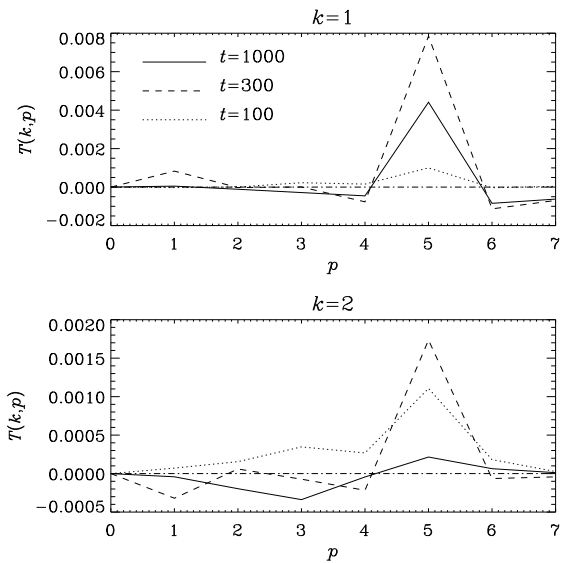}\caption[]{
Spectral energy transfer function $T(k,p,t)$, normalized by
$\bra{\BB^2}$ for three different times, for $k=1$ and 2. Run~3.
}\label{FpTM_spec}\end{figure}

In \Fig{FpTM_spec} we show $T(k,p)=\sum_q T(k,p,q)$, normalized by 
$\bra{\BB^2}$ for the corresponding times, for $k=1$ and 2. This
function shows that most of the energy of the large scale field at
$k=1$ comes from velocity and magnetic field fluctuations at the
forcing scale, $k=5$. At early times this is also true of the energy
of the magnetic field at $k=2$, but at late times, $t=1000$, the
gain from the forcing scale, $k=5$, has diminished, and instead
there is now a net loss of energy into the next larger scale, $k=3$,
suggestive of a direct cascade operating at $k=2$, and similarly at
$k=3$.

The generation of large scale energy at $k=1$ through {\it nonlocal}
inverse energy transfer is characteristic of the $\alpha$-effect
in mean-field electrodynamics. In the following we shall pursue this
analogy further. It should be emphasized, however, that without the
simultaneous loss of energy at the next smaller scales (here $k=2$ and 3)
through direct energy transfer the $k=1$ field would have been totally
swamped by smaller scale fields. Thus, nonlinearity is quite crucial
for this process to produce well defined large scale fields. Indeed,
in the absence of the nonlinear term $\JJ\times\BB$ in \Eq{dudt} the
marked large scale pattern (\Fig{Fpimages}) disappears within a
turnover time. Recent numerical experiments have shown, however,
that the ambipolar diffusion nonlinearity too leads to well defined large
scale fields -- even in the absence of the Lorentz force (Brandenburg \&
Subramanian 2000).

\subsection{Mean-field interpretation}
\label{mf_interpret}

In this subsection we adopt the hypothesis that the large scale
component of the field at wavenumber $k=1$ can be described in terms of
mean-field theory. The magnetic field at other wavenumbers is ($k\ge2$)
is important for contributing to the $\alpha$-effect and the turbulent
magnetic diffusivity, $\eta_{\rm t}$, but apart from that it is merely
an extra source of noise as far as the dynamics of the large scale field
is concerned. As we have seen in the previous section, this extra noise
is automatically kept to a minimum due to direct cascade effects and
transfer to kinetic energy during the saturation phase.

According to mean-field theory for non-mirror symmetric isotropic
homogeneous turbulence with no mean flow the mean magnetic field is
governed by the equation
\EQ
{\partial\over\partial t}\meanBB=\alpha\nab\times\meanBB
+\eta_{\rm T}\nabla^2\meanBB,
\label{dyneq}
\EN
where bars denote the mean fields, $\alpha$ and $\eta_{\rm
T}=\eta+\eta_{\rm t}$ are constants, and $\eta_{\rm t}$ is the turbulent
magnetic diffusivity. In general, these coefficients are not constant
and depend for example on the magnetic field. (In our particular
case the local magnetic energy density is however approximately
uniform.) Furthermore, since the magnetic field is strong, $\alpha$
and $\eta_{\rm t}$ should really be replaced by tensors, but we
shall ignore this additional modification except that we shall allow
$\alpha$ and $\eta_{\rm t}$ to vary slowly in time as the magnetic
field approaches saturation. This simplified form of nonlinearity may be
justified by noting that the mean magnetic field looks nearly sinusoidal
(\Fig{Frmean1000}).

Equation \eq{dyneq} permits steady force-free solutions where the current
helicity of the large scale field, $\meanJJ\cdot\meanBB$ is given by
$(\eta_{\rm T}/\alpha)\meanJJ^2$. Apart from some common phase factor,
the mean field depicted in \Fig{Frmean1000} is given by $\meanBB(y)=(\sin
y,0,\cos y)$, so $\meanJJ=(-\sin y,0,-\cos y)$, corresponding to negative
helicity, and therefore $\alpha$ must be negative. This is in agreement
with mean-field theory which predicts that $\alpha$ is a negative multiple
of the residual (kinetic minus current) helicity (e.g.\ Blackman \&
Chou 1997, Field, Blackman, \& Chou 1999), which is positive in our case;
see \Tab{T1}.

If the wavevector of the large scale field is $K_y$ (as in the case
discussed above), \Eq{dyneq} becomes
\EQ
\dot{\meanBB}_x=+\alpha\meanBB_z'+\eta_{\rm T}\meanBB_x'',
\EN
\EQ
\dot{\meanBB}_z=-\alpha\meanBB_x'+\eta_{\rm T}\meanBB_z'',
\EN
where dots and primes denote differentiation with respect to $t$ and $y$,
respectively. Since $\alpha<0$, the solution can be written in the form
\EQ
\meanBB_x(y,t)=b_x(t)\sin(y+\phi),
\EN
\EQ
\meanBB_z(y,t)=b_z(t)\cos(y+\phi),
\EN
where $b_x(t)$ and $b_z(t)$ are positive functions of time that satisfy
\EQ
\dot{b}_x=|\alpha| b_z-\eta_{\rm T} b_x,
\label{mf1}
\EN
\EQ
\dot{b}_z=|\alpha| b_x-\eta_{\rm T} b_z.
\label{mf2}
\EN
In a steady state $|\alpha|=\eta_{\rm T}$, and $b_x=b_z$. In order to
find the actual values of $|\alpha|$ and $\eta_{\rm T}$ during both
the saturated steady state and the growth phase we can do a simple
experiment: suppose we put $b_x=0$ at some moment in time, then \Eq{mf1}
would predict that $b_x$ starts to recover at the rate $|\alpha| b_z$,
which allows us to estimate $\alpha$. In practice we put $b_x=0$ by
subtracting ${\overline B}_x$ from the $x$ component of $\BB$ at a
certain time and restart the simulation with that field.

\begin{figure}[h!]\plotone{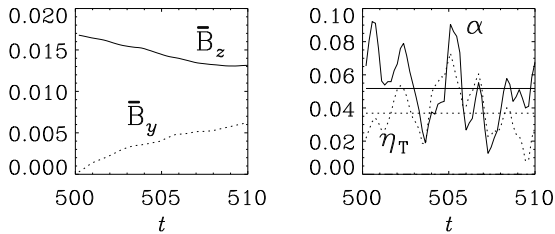}\caption[]{
Example of the evolution of the mean fields, ${\overline B}_y$ and
${\overline B}_z$, after subtracting ${\overline B}_y$ from the
actual field $B_y$ (left), and the corresponding results for $\alpha$
and $\eta_{\rm T}$ (right). The average values are indicated by
horizontal lines. Run~5.
}\label{Fpalp1b}\end{figure}

\begin{figure}[h!]\plotone{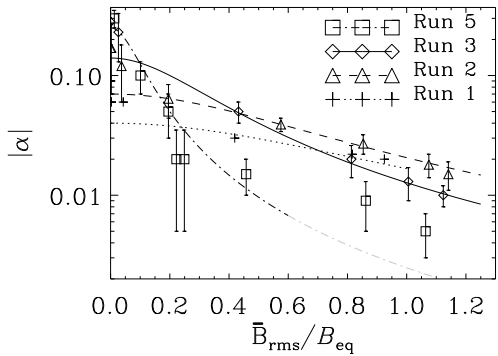}\caption[]{
Results for $\alpha$ for different values of $R_{\rm m}$ using \Eq{fit1}.
The lines represent the fits described in the text.
}\label{Fpalpb_fit2}\end{figure}

In order to have a somewhat more precise estimate we need the
solution to \eq{mf1} and \eq{mf2} for the initial condition $b_x(0)=0$;
\EQ
b_x=e^{(|\alpha|-\eta_{\rm T})t}-e^{-(|\alpha|+\eta_{\rm T})t},
\label{sol1}
\EN
\EQ
b_z=e^{(|\alpha|-\eta_{\rm T})t}+e^{-(|\alpha|+\eta_{\rm T})t},
\label{sol2}
\EN
where the amplitude is arbitrary in linear theory.
Adding and subtracting \eq{sol1} and \eq{sol2} we can solve for
$|\alpha|-\eta_{\rm T}$ and $|\alpha|+\eta_{\rm T}$, respectively. In terms
of $|\alpha|$ and $\eta_{\rm T}$ separately, we have
\EQ
|\alpha|=\half{\dd\over\dd t}\left[+\ln(+b_x+b_z)-\ln(-b_x+b_z)\right],
\label{fit1}
\EN
\EQ
\eta_T=\half{\dd\over\dd t}\left[-\ln(+b_x+b_z)-\ln(-b_x+b_z)\right].
\label{fit2}
\EN
In practice we average the results of \eq{fit1} and \eq{fit2} over
some 5--10 time units. We have applied this method to Runs~1--3 at
times between $t=100-600$ and to Run~5 at times between $t=300-1600$
(\Fig{Fpalp1b}). During these times the mean field was still evolving
(\Fig{Fpbmean}), so at different times the mean magnetic field was
different, which allows us to obtain the $\alpha(|\BB|)$ dependence.
We take into account the fact that during the experiment the actual field
is only $1/\sqrt{2}$ of what it was before one of the two components of
the mean field has been removed. We have then attempted a fit of the form
\EQ
\alpha={\alpha_0\over1+\alpha_{\rm B}\bra{\meanBB^2}/B_{\rm eq}^2},
\label{fit}
\EN
where $B_{\rm eq}^2=\mu_0\rho_0\bra{\uu^2}_{\rm sat}$. The result is shown
in \Fig{Fpalpb_fit2} and the coefficients $\alpha_0$ and $\alpha_{\rm B}$
are listed in \Tab{T2}. One should note, however, that \Eq{fit} does not
accurately represent the actual data of Run~5. Nevertheless, it is clear
that $\alpha$-quenching is enhanced for large values of $R_{\rm m}$, which
may be described by a fit of the form
\EQ
\alpha_{\rm B}\approx(R_{\rm m,forc}/22)^{1.35}.
\label{quench}
\EN
Such a steep dependence of $\alpha_{\rm B}$ on $R_{\rm m,forc}$ was
suggested by Vainshtein \ea (1993), although his argument (see also
Vainshtein \& Cattaneo 1992) assumes the presence of strong small scale
fluctuations (which is not the case here; see \Fig{Fpq}). 

\begin{figure}[h!]\plotone{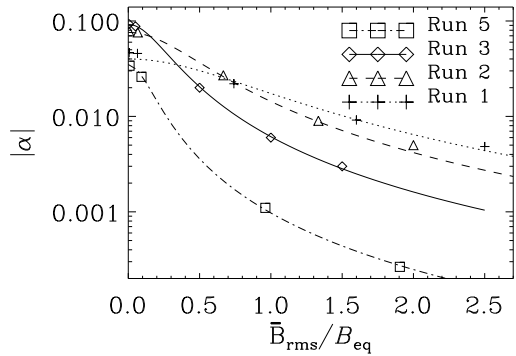}\caption[]{
Results for $\alpha$ for different values of $R_{\rm m}$ using an imposed
magnetic field. The lines represent the fits described in the text.
}\label{Fpalpb_fit2imp}\end{figure}

In \Fig{Fpalpb_fit2imp} we compare with the result for
$\alpha$ obtained by just imposing a uniform magnetic field, $B_0\zzz$,
and calculating $\alpha$ simply as
\EQ
\alpha^{\rm imp}=(\bra{\uu\times\BB})_z/B_0.
\EN
Each point in \Fig{Fpalpb_fit2imp} corresponds to a different run with
given field strength $B_0$, but otherwise the same parameters as in
Runs 1--3 and 5.  This method was frequently used in the past (e.g.\
Brandenburg \ea 1990, Tao \ea 1993, Cattaneo \& Hughes 1996), but it
is not {\it a priori} clear that one measures the same $\alpha$ as with
the method explained above. Nevertheless, the two results appear to be
qualitatively similar (cf.\ \Figs{Fpalpb_fit2}{Fpalpb_fit2imp}) although
there are some differences in the case where $R_{\rm m}$ is very large
(\Tab{T2}). The same values of $\alpha_{\rm B}$ are confirmed by yet
another method that is explained below in \Sec{SRe_dep}.

\begin{table}[h!]
\caption{Coefficients for the $\alpha$-quenching expressions
for 4 runs with different values of $R_{\rm m}$.
}\label{T2}
\begin{tabular}{lcccc}
\hline
                             & Run~1 & Run~2 & Run~3  & Run~5 \\
\hline
$R_{\rm m,lin}$              &  100  &  300  &  900   &  3600 \\
$R_{\rm m,forc}$             &   20  &   60  &  180   &   700 \\
\hline
$|\alpha_0|$                 &  0.04 &  0.07 &  0.14  &  0.30 \\
$\alpha_{\rm B}$             &  1.4  &  2.4  &   10   &  100  \\
\hline
$|\alpha_0^{\rm imp}|$       & 0.040 & 0.076 &  0.092 & 0.035 \\
$\alpha_{\rm B}^{\rm imp}$   &  1.3  &  4.3  &   14   &   35  \\
\end{tabular}
\end{table}

\begin{figure}[h!]\plotone{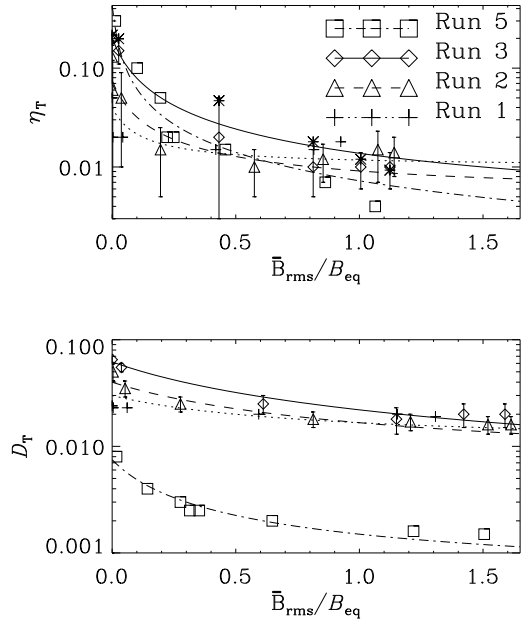}\caption[]{
Results for $\eta_T$ and $D_T$ for different values of $R_{\rm m}$.
The lines represent the fits described in the text. In the plot of
$\eta_T$ the asterisks denote $|\alpha|-\lambda$ for the
$R_{\rm m,forc}=120$ run, which agrees reasonably well with $\eta_T$.
}\label{Fpalpb_sep2}\end{figure}

The only way a strongly $R_{\rm m}$-dependent $\alpha$ quenching can
be compatible with the large scale field generation seen in the present
simulations would be if $\eta_{\rm t}$ was also strongly quenched. [See
Cattaneo \& Vainshtein (1991) for two-dimensional simulations supporting
the hypothesis of strong $\eta_{\rm t}$ quenching.] In \Fig{Fpalpb_sep2}
we compare the results obtained for $\eta_{\rm T}$ with the turbulent
diffusion coefficient for a passive scalar. The passive scalar diffusion
coefficient is obtained by simultaneously solving an additional equation
for the concentration per unit mass, $c$,
\EQ
{\DD c\over\DD t}={1\over\rho}\nab\cdot\rho D\nab c,
\EN
where $D=\eta$ is chosen for Runs~1--3 and $D=2.5\eta$ for Run~5.
The total (turbulent plus microscopic) passive scalar
diffusion coefficient is obtained by measuring the rate at which a narrow
gaussian distribution of $c$ widens as time goes on. The result is shown
in the second panel of \Fig{Fpalpb_sep2}. Generally the suppression
of $\eta_{\rm t}$ by magnetic fields is stronger than the suppression
of $D_{\rm t}$. (For Run~3 $D_{\rm T}/D$ ranges between 40 for weak
fields and 10 for strong fields.) We note that a dependence of the form
\EQ
D_{\rm T}=D+D_{\rm t0}/(1+D_{\rm B}\bra{|\meanBB|}/B_{\rm eq})
\label{Dfit}
\EN
seems to fit at least the $D_{\rm T}$ data better than a quadratic
dependence. Unlike the results for $\alpha$, the dependence of $D_{\rm
B}$ on Reynolds number is here less strong. In the fits shown in
\Fig{Fpalpb_sep2} a good fit is $D_{\rm B}=2$ for all three runs, and
$D_0=0.02$, 0.035, and 0.06 for Runs~1, 2, and 3, respectively. Run~5
behaves differently, because in this large magnetic Prandtl number run
there is no inertial range, and so we used $D_0=0.007$ and $D_{\rm B}=6$.

Determining $\eta_{\rm t}$ from dynamo simulations is notoriously difficult:
$\eta_{\rm t}$ has to be determined simultaneously with $\alpha$ from the
electromotive force, where $\eta_{\rm t}$ multiplies a derivative
of the field, so it is numerically more noisy. Nevertheless, one sees
from \Fig{Fpalpb_sep2} that $\eta_{\rm t}$ is quenched by more than
a factor of ten. However, the functional form cannot be established
from our data. Using for $\eta_{\rm T}$ a similar fit formula as \Eq{Dfit} we have
$\eta_{\rm B}=16$ for all three runs, and $\eta_{\rm T0}=0.03$, 0.07,
and 0.2 for Runs~1, 2, and 3, respectively, and $\eta_{\rm T0}=0.43$,
$\eta_{\rm B}=60$ for Run~5. However, for strong magnetic fields,
$\eta_{\rm T}$ levels off at a value of 0.01 (for Run~3) or 0.005
(for Run~5). These values are similar to the values of $\alpha$. Thus,
$\alpha\approx\eta_{\rm T}k_1$, which means that the mean-field dynamo
has turned into a marginally critical state, which is indeed to be
expected. As a consistency check for the directly obtained values of
$\eta_{\rm T}$ we use the time dependent growth rate $\lambda(t)$ and show
$|\alpha|-\lambda$ (asterisks in \Fig{Fpalpb_sep2}) for Run~3. They agree
reasonably well with $\eta_T$ (except near $\bra{|\meanBB|}=0.4B_{\rm
eq}$), suggesting that \eq{mf1} and \eq{mf2} are approximately satisfied
with the coefficients obtained above. Later in \Sec{SRe_dep} we present
a more accurate and self-consistent determination of the combined
expression for $\alpha$ and $\eta_{\rm t}$-quenching by fitting solutions
of \Eq{dyneq} to the actual evolution of the mean field. Those results
support a quadratic quenching formula for both $\alpha$ and $\eta_{\rm t}$.

\subsection{Large scale separation}
\label{Sseparation}

In some earlier exploratory models we forced the flow at $k=10$, which
gave somewhat more room for the inverse cascade to develop, but less
room for the direct cascade toward smaller wavenumbers. The latter means
that the Reynolds number with respect to the forcing scale is smaller
and the turbulent mixing properties, as quantified by the ratio $D_{\rm
t}/D$, are poorer. Strong turbulent mixing was important in order to
address the issue of Reynolds number dependent suppression of transport
coefficients. In this subsection we shall accept poor mixing in favor of
larger scale separation and hence a significantly larger subrange in which
the inverse cascade can develop. Thus, we now force the flow at $k=30$
(Run~6).

\begin{figure}[h!]\plotone{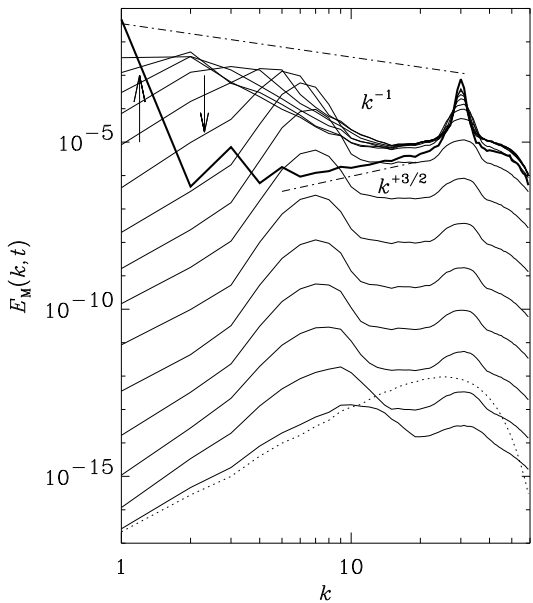}\caption[]{
Magnetic energy spectra for Run~6 with forcing at $k=30$. The times
range from 0 (dotted line) to 10, 30, ..., 290 (solid lines). The thick
solid line gives the final state at $t=1000$. Note that at early times
the spectra peaks at $k_{\max}\approx7$. The $k^{-1}$ and $k^{+3/2}$
slopes are given for orientation as dash-dotted lines.
}\label{FpMkt2}\end{figure}

After saturation is reached, which happens first at some intermediate
scale, there is a wave of spectral energy propagating toward smaller
$k$. This is similar to the results of closure models of Pouquet \ea
(1976). However, unlike these models there does not seem to develop
a $k^{-1}$ magnetic energy spectrum. Instead, there is only an {\it
envelope} of the helicity wave that follows approximately a $k^{-1}$
powerlaw (\Fig{FpMkt2}). As before, there is a strong built-up of magnetic
energy at the largest scale, $k=1$, combined with a subsequent suppression
of energy at intermediate scales ($2\leq k\leq 20$). The ratio of the peak
value at $k_1=1$ to the value of the secondary peak at the forcing scale
$k_{\rm f}=30$ scales like $(k_1/k_{\rm f})^{-1}$ (\Fig{FpMkt2}). This
is also the case in runs with $k_{\rm f}=5$ (\Fig{Fpspec_all}).

\begin{figure}[h!]\plotone{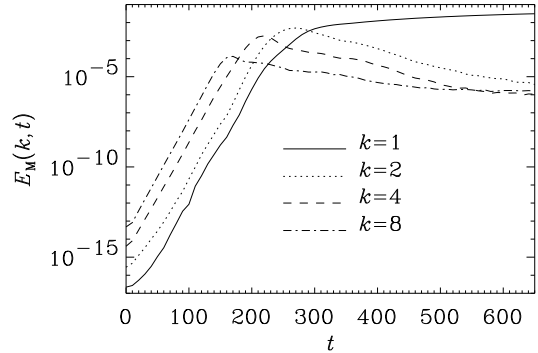}\caption[]{
Evolution of spectral magnetic energy for selected wavenumbers
in a simulation with forcing at $k=30$.
}\label{FpMkt}\end{figure}

As a function of time, the spectral magnetic energy grows at the same rate
at all values of $k$ until saturation is reached. During early times the
magnetic spectra peak at $k_{\max}\approx7$, which is also the wavenumber
that reaches equipartition first, but then the field at this wavenumber
decays to a somewhat smaller value while the contributions from the
next smaller $k$ grow and subsequently decay; see \Fig{FpMkt}. Finally,
when the energy at $k=1$ reaches equipartition the energy of all larger
$k$s becomes suppressed.

Although a comparison with mean-field theory may be inappropriate, it is
interesting to note that the existence of a $k_{\max}$ is predicted from
the theory of $\alpha^2$ dynamos in infinite media (Moffatt 1978). The
dispersion relation is
\EQ
\lambda=|\alpha|k-\eta_{\rm T}k^2,
\EN
where $\lambda$ is the growth rate. Maximum growth occurs at wavenumber
\EQ
k_{\max}=\half|\alpha|/\eta_{\rm T},
\EN
and there the maximum growth rate is
\EQ
\lambda_{\max}=\half|\alpha|k_{\max}.
\EN
Since $k_{\max}$ and $\lambda_{\max}$ can be measured, we
can determine
\EQ
|\alpha|=2\lambda_{\max}/|k_{\max},\quad
\eta_{\rm T}=\lambda_{\max}/|k_{\max}^2
\EN
during the growth phase of the dynamo. With $k_{\max}=7$ (\Fig{FpMkt})
and $\lambda_{\max}=0.07$ (\Fig{FpMkt2}) we have $|\alpha|=0.02$ and
$\eta_{\rm T}=0.0014$. In this case, the forcing occurs essentially in the
dissipation range, so the turbulence has therefore poor mixing properties
($\eta_{\rm T}/\eta=1.4$). The value of $\alpha$ is about 7 times smaller
than that of Run~3 during the kinematic regime, which seems reasonable.

In the nonlinear regime there are marked differences between mean-field
theory and simulations. Using a nonlinear mean-field model of the inverse
cascade with $\alpha$-effect, Galanti, Sulem, \& Gilbert (1991) found
significant power at $k\ge2$, which is in contrast to the behavior seen
in \Fig{FpMkt2}. Consequently, the mean fields of Galanti, Gilbert,
\& Sulem (1990) look much less sinusoidal than in the simulations
(\Fig{Frmean1000}). If one wanted to model this within the framework of
an $\alpha$-effect one would need to invoke a scale-dependent $\alpha$
integral kernel, where the dominant contributions to $\alpha$ come only
from the largest scale of the system (Brandenburg \& Sokoloff 2000).

\subsection{Reynolds number dependence and magnetic helicity}
\label{SRe_dep}

In this section we first discuss to which extent our results are affected
by the limited resolution and finite magnetic Reynolds number. We
then show that magnetic helicity conservation implies slow saturation
of helical magnetic fields, and that this features is quantitatively
reproduced by a mean-field model with $R_{\rm m}$-dependent quenching
of $\alpha$ and $\eta_t$.

\begin{figure}[h!]\plotone{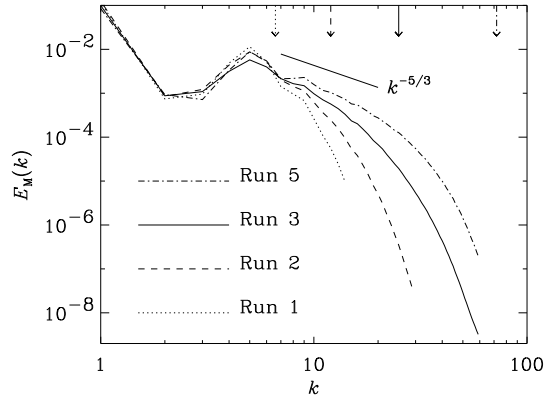}\caption[]{
Comparison of time averaged magnetic energy spectra for Runs~1--3
($t=600-1000$) and Run~5 ($t=1600$). To compensate for different field
strengths and to make the spectra overlap at large scales, two of the
three spectra have been multiplied by a scaling factor ($\times3.4$ for
Run~5, $\times2$ for
Run~2 and $\times5$ for Run~1). There are signs of a gradual development
of an inertial subrange for wavenumbers larger than the forcing scale. The
$k^{-5/3}$ slope is shown for orientation. The dissipative magnetic
cutoff wavenumbers, $k_{\rm d}=\bra{\JJ^2/\eta^2}^{1/4}$, are indicated
by arrows at the top.
}\label{Fpspec_conv}\end{figure}

In \Fig{Fpspec_conv} we show energy spectra of the magnetic field for
three values of $R_{\rm m}$. The three curves differ by a certain factor,
but are otherwise essentially the same at small wavenumbers. The main
difference is, as expected, at the smaller scales: the spectra for large
$R_{\rm m}$ begin to show signs of an inertial range for values of $k$
larger than the forcing wavenumber.

Although the magnetic energy spectra of the statistically steady
state seem to show convergence to a spectrum roughly compatible
with $k^{-5/3}$, there are some serious concerns about the {\it
timescale} on which such a steady state is achieved.  As was
first pointed out by Berger (1984), in a closed box (periodic or
perfectly conducting) there is an upper bound on the rate of change
of the magnetic helicity.  This is relevant, because the fields
that are generated in the simulations have strong magnetic helicity:
$\bra{\AAA\cdot\BB}/\bra{\BB^2}\approx-0.8/k_1$ (\Fig{Fpcorrel}). Most
of the magnetic helicity is in the large scales (\Fig{Fpspec_all}). In
order to build up the helicity at large scales we have to destroy magnetic
helicity at small scales (\Fig{Fpspec_helic2}).

Open boundary conditions would help to get rid of magnetic helicity (cf.\
Blackman \& Field 2000), which may well be an important factor in more
realistic simulations (e.g.\ Glatzmaier \& Roberts 1995, Brandenburg
\ea 1995).  However, the present simulations are for closed boxes and
yet they do show strong fields, so we need to understand whether and how
they have been affected by this constraint.

The rate at which magnetic helicity can change, see \Eq{HelCons}, is
constrained by the Schwarz inequality,
\EQ
\left|{\dd\over\dd t}\bra{\AAA\cdot\BB}\right|=2\eta\,|\bra{\JJ\cdot\BB}|
\leq 2\eta\, J_{\rm rms} B_{\rm rms}
\EN
(Berger 1984, Moffatt \& Proctor 1985). In forced systems it is
common (albeit not necessary) that the energy ($B_{\rm rms}^2$)
and dissipation rate ($\eta J_{\rm rms}^2$) are independent of the
Reynolds number. However, this would imply that the rate of change of
$\bra{\AAA\cdot\BB}$ is limited resistively by a term proportional to
$\eta^{1/2}$. In our simulations, $\bra{\JJ\cdot\BB}/(J_{\rm rms}B_{\rm
rms})$ has a maximum of $\sim0.25$ (see \Tab{T1}), so this limit does
not seem to be saturated. Another estimate for the helicity dissipation
is obtained by assuming that the magnetic spectrum has powerlaw behavior
with $E_{\rm M}(k)\sim k^{-n}$ for $k<k_{\rm d}$. The value of $k_{\max}$
follows by assuming that the Joule dissipation is independent of
$\eta$, which yields $k_{\rm d}\sim\eta^{-1/(3-n)}$ ($\sim\eta^{-3/4}$
for $n=5/3$). The helicity dissipation is then proportional to $k_{\rm
d}^{-1}\sim\eta^{1/(3-n)}$, which always tends to zero for small values
of $\eta$. For $n=5/3$ the magnetic helicity dissipation scales like
$\eta^{3/4}$, which is faster than Berger's limit.

There is a related and even more important consequence of helicity
conservation. In a steady state, $\bra{\AAA\cdot\BB}$, which is gauge
invariant and therefore physically meaningful, must be constant, and
hence $\bra{\JJ\cdot\BB}\rightarrow0$. Thus, although we are going to
generate a strong large scale field with significant current helicity in
the large scales, the net current helicity must actually vanish. This can
only happen if there is an equal amount of small scale current helicity
of opposite sign, i.e.\
\EQ
\bra{\jj\cdot\bb}\approx-\bra{\meanJJ\cdot\meanBB}
\quad\mbox{(in the steady state)},
\EN
where angular brackets denote volume averages, bars denote the
large scales at $k=1$, and lower characters refer to contributions
from all higher wavenumbers. The spectrum of the magnetic helicity
is $k^2$ times steeper than that of the current helicity, so
$\bra{\aaa\cdot\bb}\approx\bra{\jj\cdot\bb}/k_{\rm f}^2$, and
therefore $|\bra{\aaa\cdot\bb}|\ll|\bra{\meanAA\cdot\meanBB}|$.

In the following we estimate the evolution of the energy density of
the large scale field, $\meanBB^2$, using the fact that in the present
simulations a large fraction of the magnetic energy is contained in the
large scale field (\Fig{Fpq}) and that magnetic helicity is strong
(\Fig{Fpcorrel}). This would be too ideal an assumption for
astrophysical settings, but it is adequate for describing our present
simulations. Thus, we may set
\EQ
k_1\bra{\AAA\cdot\BB}
\approx k_1\bra{\meanAA\cdot\meanBB}
\approx-\bra{\meanBB^2}
\approx\bra{\meanJJ\cdot\meanBB}/k_1.
\EN
For clarity we retain here the $k_1$ factors, even though they are
one. Using \Eq{HelCons} this yields
\EQ
{\dd\over\dd t}\bra{\meanBB^2}\approx-2\eta k_1^2\bra{\meanBB^2}
+2\eta k_1|\bra{\jj\cdot\bb}|,
\label{evolBB2}
\EN
which has the solution
\EQ
\bra{\meanBB^2}
\approx2\eta k_1 e^{-2\eta k_1^2t}\int_{t_{\rm sat}}^te^{2\eta k_1^2t'}
|\bra{\jj\cdot\bb}|\,\dd t',
\label{solBB2}
\EN
which is expected to apply after the time of saturation, $t=t_{\rm sat}$,
when $\bra{\jj\cdot\bb}$ is approximately constant, so
\EQ
\bra{\meanBB^2}\approx{|\bra{\jj\cdot\bb}|\over k_1}
\left[1-e^{-2\eta k_1^2(t-t_{\rm sat})}\right].
\label{approxB}
\EN
To a good approximation we may assume
$\bra{\jj\cdot\bb}\approx\rho_0\bra{\oo\cdot\uu}$. This
means that the `residual helicity' (Pouquet \ea 1976),
$\bra{\oo\cdot\uu}-\bra{\jj\cdot\bb}/\rho_0$, is small, which is
indeed consistent with the present data and also with a results of
Brandenburg \& Subramanian (2000), who used the ambipolar diffusion
model of Subramanian (1999). The kinetic helicity can be approximated
by $\bra{\oo\cdot\uu}\approx k_{\rm f}\bra{\uu^2}$, and so the final
field strength, $B_{\rm fin}$, is given by
\EQ
B_{\rm fin}^2
\approx{|\bra{\jj\cdot\bb}|\over k_1}
\approx\rho_0{|\bra{\oo\cdot\uu}|\over k_1}
\approx\rho_0\bra{\uu^2}{k_{\rm f}\over k_1}
\equiv B_{\rm eq}^2{k_{\rm f}\over k_1},
\label{approxBfin}
\EN
For Runs~1--3 and 5--7, the ratio of the actual values of $B_{\rm fin}$
to those anticipated from \Eq{approxBfin} are 0.61, 0.76, 0.85, 1.16,
0.74, and 0.36. In the large $R_{\rm m}$ cases (Runs~3 and 5) the ratio
is close to one, confirming thus the assumption of small residual helicity
in the saturated state.

\begin{figure}[h]\plotone{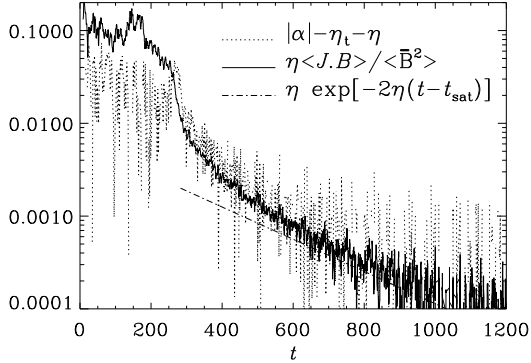}\caption[]{
Comparison of $|\alpha|-\eta_{\rm T}k_1$ with
$\eta\bra{\JJ\cdot\BB}/\bra{\meanBB^2}$ and $\eta\exp[-2\eta
k_1^2(t-t_{\rm sat})]$. Note that $|\alpha|-\eta_{\rm T}k_1$ follows
very closely the resistively dominated limit. Notice also the relatively
large noise level in $|\alpha|-\eta_{\rm T}$. Run~3.
}\label{Fpalp_corr}\end{figure}

The resistively limited growth of $\meanBB$ has immediate consequences
for $\alpha$. The evolution equation for $\bra{\meanAA\cdot\meanBB}$
can be derived from \Eq{dyneq},
\EQ
{\dd\over\dd t}\bra{\meanAA\cdot\meanBB}
=2\alpha\bra{\meanBB^2}-2\eta_{\rm T}\bra{\meanJJ\cdot\meanBB}.
\label{heldyneq}
\EN
In the strongly helical case considered here the magnetic helicity
of the large scale field, $\bra{\meanAA\cdot\meanBB}$, is very nearly
equal to $\bra{\AAA\cdot\BB}$, so the right hand sides of \eq{HelCons}
and \eq{heldyneq} must be approximately equal, which leads to
\EQ
|\alpha|-\eta_{\rm T}k_1\approx\eta|\bra{\JJ\cdot\BB}|/\bra{\meanBB^2}.
\label{alpconstr}
\EN
In order to check this relation we plot the two sides of \eq{alpconstr}
versus time (\Fig{Fpalp_corr}), where $\eta_{\rm T}=\eta+\eta_{\rm t}$
and $|\alpha|-\eta_{\rm t}k_1$
has been obtained directly from $\emf=\uu\times\BB$ assuming
$\meanEE=\alpha\meanBB-\eta_{\rm t}\meanJJ$ which, for fully helical
mean fields, leads to
\EQ
|\alpha|-\eta_{\rm t}k_1=|\bra{\meanEE\cdot\meanBB}|/\bra{\meanBB^2}.
\EN
(Note that here $\eta_{\rm t}$ enters, not $\eta_{\rm T}$. Recall also that in
our case $\alpha<0$, so $\bra{\meanEE\cdot\meanBB}<0$.) The condition
\eq{alpconstr} on $\alpha$ reflects the fact that the growth of the large
scale field is limited by the microscopic resistivity, as seen already
from \Eq{approxB}. Note, however, that this statement only applies to
the present case of strong helicity and closed or periodic boxes.

\begin{figure}[h!]\plotone{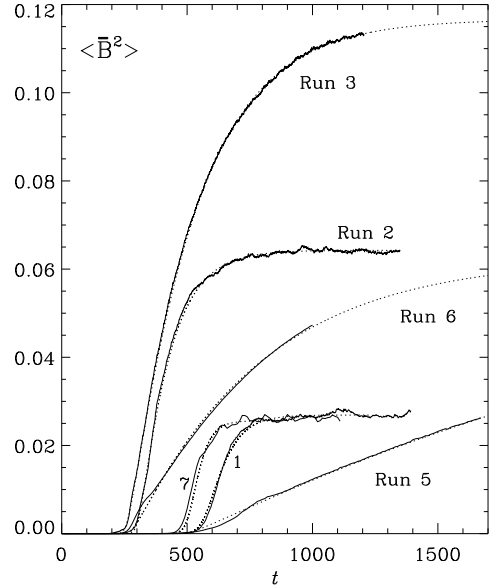}\caption[]{
Evolution of $\bra{\meanBB^2}$ for Runs~1--3 and 5--7 (solid lines), compared
with the solution \eq{approx_quench} of the mean-field dynamo equations using
resistively dominated $\alpha$ and $\eta_{\rm t}$ quenchings (dotted lines)
}\label{Fpjbm_decay_nfit}\end{figure}

There have been previous attempts to incorporate the conservation of
helicity into the expression for $\alpha$, which led to a dynamical
dependence of $\alpha$ on time and field strength (Kleeorin, Rogachevskii,
\& Ruzmaikin 1995, Kleeorin \ea 2000). In the following we point out,
however, that the helicity constraint, which leads to the prolonged saturation
phase, is well described in terms of an $\alpha^2$-dynamo with resistively
dominated quenching functions for both $\alpha$ and $\eta_{\rm t}$, i.e.\
\EQ
\alpha={\alpha_0\over1+\alpha_B\meanBB^2\!/B_{\rm eq}^2},\quad
\eta_{\rm t}={\eta_{\rm t0}\over1+\eta_B\meanBB^2\!/B_{\rm eq}^2},
\label{quench_both}
\EN
where $\alpha_B=\eta_B$ is assumed. Using the fact that the magnetic
energy density of the mean field, $\meanBB^2$, is approximately uniform
we can write \Eq{dyneq} in the form
\EQ
{\dd\over\dd t}\left(\half\meanBB^2\right)={|\alpha_0|k_1-\eta_{\rm t0}k_1^2
\over1+\alpha_B\meanBB^2\!/B_{\rm eq}^2}\meanBB^2-\eta k_1^2\meanBB^2.
\label{timedependent}
\EN
There is a steady solution with
\EQ
\alpha_B\meanBB^2/B_{\rm eq}^2=\lambda/\eta k_1^2,
\label{final}
\EN
where
\EQ
\lambda=|\alpha_0|k_1-\eta_{\rm T0}k_1^2
\EN
is the kinematic growth rate of the dynamo and $\eta_{\rm
T0}=\eta+\eta_{\rm t0}$. The time dependent equation \eq{timedependent}
can be integrated to give the solution in the form
\EQ
\meanBB^2/(1-\meanBB^2\!/B_{\rm fin}^2)^{1+\lambda/\eta k_1^2}
=B_{\rm ini}^2\,e^{2\lambda t},
\label{approx_quench}
\EN
where $B_{\rm ini}$ is the initial field strength and $B_{\rm fin}$ is
the final field strength which is given by \Eq{final}. The parameters
$B_{\rm ini}$ and $B_{\rm fin}$ can be obtained from \Fig{Fpenerg}. The
kinematic growth rate $\lambda$ is the same for small and large scale
fields and can hence be taken from \Tab{T1}. We emphasize that there
is excellent agreement between the results of the simulation and the
solution \eq{approx_quench}; see \Fig{Fpjbm_decay_nfit}.

Having obtained $\lambda$ and $B_{\rm fin}$ from the simulation
we can use \Eq{final} to find for $\alpha_B$ and $\eta_B$ the result
\EQ
\alpha_B=\eta_B={\lambda\over\eta k_1^2}
\left({B_{\rm eq}\over B_{\rm fin}}\right)^2.
\label{alpcoef}
\EN
Note that $\alpha_B$ and $\eta_B$ are proportional to $1/\eta$, which
implies $R_{\rm m}$-dependent $\alpha$ and $\eta_{\rm t}$ quenchings. The
results for $\alpha_B=\eta_B$ are summarized in \Tab{T3} for different
runs.

To a good approximation $B_{\rm fin}^2/B_{\rm eq}^2\approx k_{\rm f}/k_1$,
see \Eq{approxBfin}, so we expect $\alpha_B\approx\lambda/(\eta k_1 k_{\rm
f})$. Since $\lambda\approx 0.3\times u_{\rm rms}/\ell_{\rm f}$
we have
\EQ
\alpha_B\approx{0.3\over(2\pi)^2}\,{u_{\rm rms}L\over\eta}
\approx0.01\times R_{\rm m,lin}.
\label{alpcoef_estim}
\EN
Thus, $\alpha_B$ should scale with the magnetic Reynolds number based on
the box scale -- not the forcing scale. The difference is particularly
evident when comparing with Run~6 where the forcing wavenumber is six
times larger than in the other runs. For comparison we list in \Tab{T3}
also $R_{\rm m,lin}/100$ and $R_{\rm m,forc}/20$. The agreement between
\Eqs{alpcoef}{alpcoef_estim} is generally good, except for Run~7 where
$\alpha_{\rm B}$ is larger than expected. This is probably because in
this run with a small value of $\nu/\eta$ and small $R_{\rm m}$ the
value of $B_{\rm fin}$ is smaller than expected from \Eq{approxBfin};
see \Eq{alpcoef}.

In \Tab{T3} we also give for completeness the values of $B_{\rm
ini}$. Together with the values of $\eta$ and $\lambda$ in \Tab{T1}
all the data entering \Eq{approx_quench} are now specified. It should
be mentioned that the value of $B_{\rm ini}$ is obtained from a fit
and is only roughly comparable to the actual seed magnetic field in the
simulation, where different initial structures are possible.

\begin{table}[h!]
\caption{Coefficients for the $\alpha$-quenching expression obtained
from \Eq{alpcoef} for runs with different values of magnetic Reynolds
number. Note the rough agreement between the values of $R_{\rm m,lin}/100$
and $\alpha_{\rm B}$ for all runs except Run~7.
}\label{T3}
\begin{tabular}{cccccl}
& ${\displaystyle{R_{\rm m,lin}\over100}}$
& ${\displaystyle{R_{\rm m,forc}\over20}}$
& $\alpha_{\rm B}$ & $B_{\rm fin}$ & $B_{\rm ini}$ \\
\hline
Run~1 &   1   &   1   &  1.4 &  0.16 & $3\times10^{-8}$  \\
Run~2 &   3   &   3   &  3.9 &  0.25 & $3\times10^{-9}$  \\
Run~3 &   9   &   9   &  9.3 &  0.34 & $2\times10^{-9}$  \\
Run~5 &  35   &  35   &  30  &  0.27 & $4\times10^{-11}$ \\
Run~6 &   5   &   1   &  4.6 &  0.25 & $4\times10^{-11}$ \\
Run~7 &   1   &   1   &  2.8 &  0.16 & $1\times10^{-9}$  \\
\hline
\end{tabular}
\end{table}

For weak fields \Eq{approx_quench} gives the usual exponential growth,
$|\meanBB|=B_{\rm ini}\exp(\lambda t)$. For strong fields we recover
\Eq{approxB} with $t_{\rm sat}=\lambda^{-1}\ln(B_{\rm fin}/B_{\rm ini})$
in the limit $\lambda\gg\eta k_1^2$.

\subsection{Large magnetic Prandtl numbers}
\label{SlargePr}

The magnetic helicity constraint becomes more important as the magnetic
Reynolds number is increased. So far we have mainly considered the case
where $\nu/\eta=1$. In the sun and many other astrophysical bodies,
$\nu/\eta\ll1$, but in the galaxy $\nu/\eta\gg1$ (e.g.\ Kulsrud \&
Andersen 1992). This may lead to a magnetic energy spectrum peaking at
small scales (Kulsrud \& Andersen 1992). However, although viscous damping
will dissipate energy at small scales (Chandran 1998, Kinney \ea 1998),
there is some recent evidence that the inverse cascade may no longer
operate (Maron 2000). These results have been obtained in
the absence of net helicity. It will therefore be interesting to see
whether in the presence of net helicity an inverse cascade is still
possible when $\nu/\eta\gg1$.

In a preliminary attempt to clarify this question we have carried out
simulations for $\nu/\eta=20$ (mainly by increasing the viscosity to
$\nu=0.02$; Run~4) and $\nu/\eta=100$ (where the magnetic
diffusivity was lowered to $\eta=2\times10^{-4}$; Run~5). The viscous
cutoff wavenumber is then around 5, i.e.\ at the forcing scale, and
the magnetic cutoff wavenumbers are 25 and 72, respectively. The resolution
for Run~5 may be insufficient and discretisation errors must play a role
at small scales, but the images of the current density look
reasonable (see below) and the evolution of the large scale field
is also in agreement with expectations (\Sec{SRe_dep}).

\begin{figure}[h!]\plotone{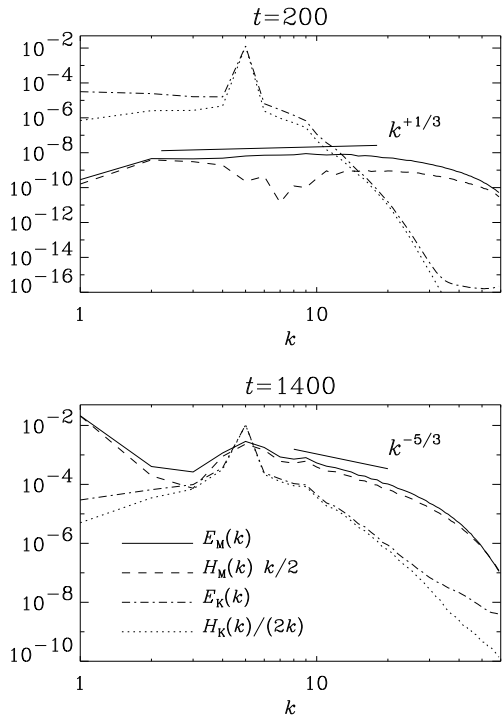}\caption[]{
Spectra of magnetic and kinetic energy and helicity for a large magnetic
Prandtl number at early and late times. $\nu=0.02$, $\eta=2\times10^{-4}$.
}\label{Fpspec_hipr}\end{figure}

\begin{figure}[h!]\plotone{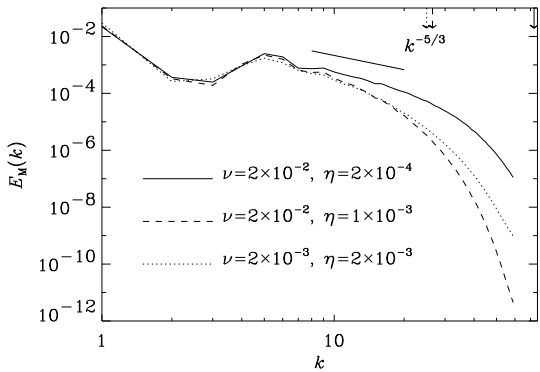}\caption[]{
Comparison of the magnetic energy spectra for Runs~3, 4 and 5. The magnetic
cutoff wavenumbers are 25, 26, and 72, as indicated by arrows at the top.
}\label{Fpspec_hipr_comp}\end{figure}

\begin{figure}[h!]\plotone{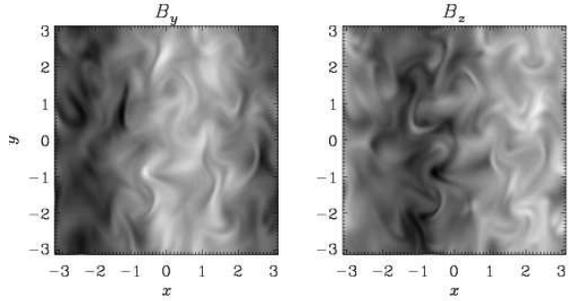}\caption[]{
Images of the $B_y$ and $B_z$ components of the magnetic field in an
arbitrarily chosen $xy$ plane. Run~5, $t=1600$.
}\label{Fbfield_KK5.n02e001d}\end{figure}

As seen from \Fig{Fpspec_hipr} the magnetic energy spectrum does
indeed peak at large wavenumbers initially, although the magnetic
energy spectrum does not scale like $k^{3/2}$ (Kulsrud \& Andersen
1992). Instead, the spectrum is close to $k^{1/3}$, which was also
found during the kinematic stage of convective dynamos (Brandenburg
\ea 1996). However, this result is inconclusive, because it could be
an artifact of the lack of an inertial range in this run. In any case,
at large wavenumbers the magnetic energy exceeds the kinetic energy,
although at later times the kinetic energy is increased somewhat by
magnetic forces. Especially at later times the magnetic energy is no
longer dominated by small scales and the spectrum falls off more like
$k^{-5/3}$. The convergence to this powerlaw is evident when comparing
Runs~3, 4, and 5 (\Fig{Fpspec_hipr_comp}). Most importantly, there are
now clear signs of an inverse cascade; see \Fig{Fbfield_KK5.n02e001d}.

\begin{figure}[h!]\plotone{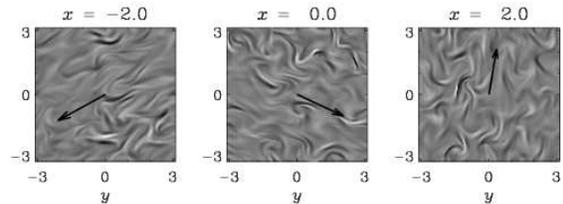}\caption[]{
Images of the current density, $J_x$, in the plane of the large scale
magnetic field. The field direction is shown as a vector. Run~5, $t=1600$.
}\label{Fbperp}\end{figure}

In these runs with large magnetic Prandtl number the current density
shows strong filamentary structures that tend to be aligned with the
local magnetic field direction, as seen in \Fig{Fbperp}. The resulting
anisotropy affects particularly the small scales (Goldreich \& Sridhar
1997, Maron 2000). Note that this type of anisotropy cannot be captured
by closure models (e.g.\ Pouquet \ea 1976).

The main shortcoming of the present large Prandtl number calculations is
that the viscous dissipation cutoff wavenumber is so small that it lies
in the range of the forcing scale, so no inertial range in the kinetic
energy is possible. At the same time, of course, the range of scales
available to the magnetic field is still not large enough to establish
a $k^{3/2}$ scaling at early times.

\section{Conclusions}

The main conclusion to be drawn from this work is that in the presence of
net magnetic helicity there is a gradual built-up of a nearly force-free
magnetic field at the largest possible scale of the system. In our
periodic calculations this corresponds to a sinusoidal one-dimensional
Beltrami field, e.g.\ $\meanBB\sim(\cos z,\sin z,0)$, which is of
course locally strongly distorted by the turbulence. Nevertheless,
the presence of the large scale field is clearly seen without averaging
(\Fig{Fpimages}). We emphasize that this result is numerically robust:
the relative dominance of magnetic energy at the smallest wavenumber
is independent of resolution (\Fig{Fpspec_conv}) and independent of
the degree of scale separation (\Fig{FpMkt2}). So, the effect is seen
equally well at resolutions ranging from $30^3$ to $120^3$ meshpoints,
and at forcing wavenumbers ranging from 5 to 30. However, the time it
takes to establish such large scale fields increases with the ohmic
diffusion time. We also note that the results are not very sensitive
to the choice of the forcing function: a forcing function that is
nearly delta-correlated in space, but still strongly helical, yields
very similar results. In the absence of net helicity, however, no large
scale field is generated. Likewise, if the forcing is made non-helical the
large scale field disappears.

An important property of the turbulence is that once the large scale
field is established, it can suppress magnetic energy on scales smaller
than the largest one. This leads to something like a `self-cleaning'
processes. This is also seen in histograms of the magnetic field which
are, for the present simulations, more nearly gaussian (with one hump
perpendicular to the field, and two in the direction of the field). This
is similar to other simulations with large scale dynamo action (see
Brandenburg \ea 1995), but
very different from simulations of small scale dynamo action where
the histograms of the field components show stretched exponentials
(Brandenburg \ea 1996), which can also be seen in the present simulations,
but only at early times.

\begin{figure}[h!]\plotone{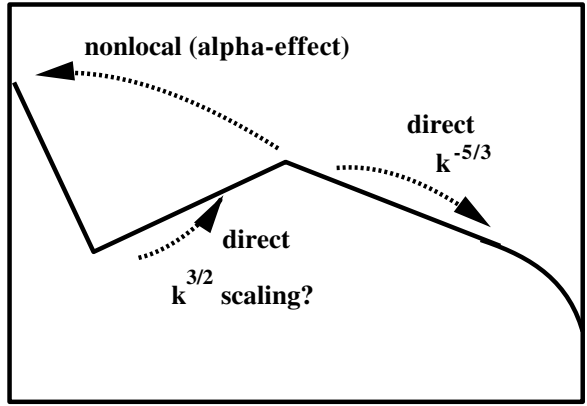}\caption[]{
Sketch illustrating direct and inverse cascade processes in helical
MHD turbulence.
}\label{Fpower_sk}\end{figure}

Our simulations show that most of the energy input to the large scale
field comes from small scales. This type of nonlocal spectral energy transfer is
suggestive of an $\alpha$-effect that could be responsible for the field
generation, rather than a local inverse cascade, which transports energy
from $k=2$ to $k=1$, for example. Although a local inverse cascade
seems to occur at early times, i.e.\ before the magnetic field is fully
established, once the field is strong the magnetic energy at $k=2$ is
actually cascaded to $k=3$ and/or transferred to kinetic energy, both of
which are probably important for the `self-cleaning' processes.  A sketch
of the anticipated energy transfer properties is given in \Fig{Fpower_sk}.

We point out that the present simulations must not be regarded as
local in the sense of representing only a small chunk of a larger
system, because the field structure depends crucially on the size of
the box. Instead, they should be viewed as global within the geometry
considered. With other boundary conditions or in different geometries
the shape of the large scale field will be different. In the case of a
sphere, for example, no perfectly force-free field is possible, but the
field may be nearly force-free. An example may be the field obtained in
hydromagnetic calculations with $\alpha$-effect (Proctor 1977), where
field saturation occurs through the Lorentz force of the large scale
field. In these calculations the magnetic saturation field strength
is relatively large, which reflects the fact that the field is indeed
nearly force-free.

It should be emphasized that the overall growth of the large scale field
and the saturation phase of the dynamo are well described by a simple
$\alpha^2$-dynamo with $\alpha$ and $\eta_{\rm t}$ coefficients that are
quenched in a $R_{\rm m}$-dependent fashion; see
\Eqs{quench_both}{alpcoef}. The reason such a dynamo can
still saturate is because of the presence of microscopic diffusion, and
it is this what causes the saturation to happen so slowly. The excellent
agreement in the evolution toward saturation between both the simulation
and the mean-field model is an indication that the simple quadratic
quenching formula is actually correct. For example a cubic nonlinearity
(Moffatt 1972, R\"udiger 1974) would lead to different behavior and
would not have the correct resistive relaxation asymptotics consistent
with helicity conservation (Brandenburg 2000).

The slow resistive field evolution past equipartition has become
particularly clear in Run~5, where the final selection of the large
scale field structure occurred rather late (after $t\approx1200$,
corresponding to about 100 turnover times; \Fig{Fpbmean}). By contrast,
in Run~3, where the magnetic Reynolds number was about 6 times smaller,
the large scale field was fully developed by the time $t\approx400$,
corresponding to about 50 turnover times. In stars the typical magnetic
Reynolds numbers are at least another six orders of magnitude larger than
in Run~5, so a large scale field, if generated by an $\alpha$-effect,
would require $\sim10^8$ turnover times or $\sim3\times10^6\yr$
(assuming a turnover time of 10 days). In the case of the sun this
estimate would be reduced by another factor of 100 (Brandenburg
\ea 2000), because differential rotation contributes to non-helical
field generation, so the resulting fields are only partially subject to
the helicity constraint. Since even the youngest protostars are older than
$3\times10^4\yr$ the $\alpha\omega$-dynamo may still be responsible for
field generation in these bodies. For galaxies, on the other hand, the
magnetic Reynolds numbers are by another seven orders of magnitude larger
than in stars, making here the case for an $\alpha\omega$-dynamo more
doubtful. However, this assumes that the conclusions from models with
closed or periodic boundaries apply to galaxies, and that the microscopic
resistivity is not enhanced during reconnection [see Ji \ea (1998)
for anomalous resistivities in a laboratory reconnection experiment].

There is now also some evidence that in oscillatory dynamos of
$\alpha\omega$-type the cycle
period is not strongly affected by the helicity timescale constraint
(Brandenburg \ea 2000). This could be related to the fact that with
shear the large scale field is no longer fully force-free and that in
that case the turbulent magnetic diffusivity is only partially suppressed
(Gruzinov \& Diamond 1996). However, the case for $\alpha\omega$-dynamo
action in stars, galaxies or accretion discs is by no means settled.
Firstly, proposals have been made for nonhelical large scale dynamo
action (Vishniac \& Cho 2000, Zheligovsky, Podvigina, \& Frisch 2000),
which may avoid the problems that $\alpha\omega$-dynamos have. Secondly,
real astrophysical bodies do have open boundaries and may get rid of small
scale helicity rather rapidly (Berger \& Ruzmaikin 2000). Indications
are, however, that open boundaries also produce significant losses at
large scales, which lowers the overall dynamo efficiency; preliminary
results are reported in Brandenburg (2000).

\acknowledgments
I thank the participants of the Astrophysical Turbulence Program at the
Institute for Theoretical Physics at the University of California, Santa
Barbara, for the many valuable discussions without which this work would
not have been possible. I am grateful for the referees' criticism which
lead to considerable improvements in the presentation of the results. This
work was supported by the National Science Foundation through the
grant PHY94-07194 and PPARC through the grant PPA/G/S/1997/00284. Use
of the PPARC supported supercomputers in St Andrews and Leicester is
acknowledged.

\end{document}